\documentclass[11pt]{article}
\usepackage{array,multirow}
\usepackage[utf8]{inputenc}
\usepackage[linesnumbered,ruled]{algorithm2e}
\usepackage{amsmath}
\usepackage{graphicx}
\usepackage{caption}
\usepackage{subcaption}
\usepackage{datetime}
\usepackage{longtable}
\usepackage{booktabs}
\usepackage{lscape}
\usepackage{hyperref}
\hypersetup{
    colorlinks = true,
    allcolors = blue
}
\usepackage{times}
\usepackage{float}
\restylefloat{table}

\usepackage[margin=1in]{geometry}

\usepackage[parfill]{parskip}
\usepackage{setspace}
\usepackage[title]{appendix}

\usepackage[dvipsnames,table]{xcolor}
\usepackage[table]{xcolor}
\usepackage{tikz}
\usetikzlibrary{backgrounds}

\usepackage{anyfontsize}

\usepackage{enotez} % for endnotes

\usepackage{amssymb}

\usepackage[inline]{enumitem}

\usepackage{amsmath}
\usepackage[round,authoryear]{natbib}
\usepackage{authblk}
\usepackage{tikz}
\usetikzlibrary{decorations.pathreplacing,positioning, arrows.meta}

\usepackage{listings}

\definecolor{mygray}{gray}{0.8}

\newcolumntype{L}[1]{>{\RaggedRight}p{#1}}

\definecolor{textcolor1}{HTML}{000000} 
\definecolor{textcolor2}{HTML}{000000} 
\begin{document}
\title{Not All Bonds Are Created Equal: Dyadic Latent Class Models for Relational Event Data}

 \author[a]{Rumana Lakdawala \footnote{\textbf{Corresponding Author:} \newline
Department of Methodology and Statistics, Tilburg School of Social and Behavioral Sciences, Tilburg University, Tilburg University, Warandelaan 2, 5037 AB Tilburg, The Netherlands \\ 
Email: r.j.lakdawala@tilburguniversity.edu}}
\author[b,c]{Roger Leenders}
\author[a]{Joris Mulder}

\affil[a]{Department of Methodology and Statistics, Tilburg School of Social and Behavioral Sciences, Tilburg University, The Netherlands}

\affil[b]{Jheronimus Academy of Data Science, The Netherlands}

\affil[c]{Department of Organization Studies, Tilburg School of Social and Behavioral Sciences, Tilburg University, The Netherlands}

\date{}

\maketitle
\noindent\rule{\textwidth}{0.5pt}
\begin{abstract}
Dynamic social networks can be conceptualized as sequences of dyadic interactions between individuals over time. The relational event model has been the workhorse to analyze such interaction sequences in empirical social network research. When addressing possible unobserved heterogeneity in the interaction mechanisms, standard approaches, such as the stochastic block model, aim to cluster the variation at the actor level. Though useful, the implied latent structure of the adjacency matrix is restrictive which may lead to biased interpretations and insights. To address this shortcoming, we introduce a more flexible dyadic latent class relational event model (DLC-REM) that captures the unobserved heterogeneity at the dyadic level. Through numerical simulations, we provide a proof of concept demonstrating that this approach is more general than latent actor-level approaches. To illustrate the applicability of the model, we apply it to a dataset of militarized interstate conflicts between countries.
\end{abstract}
\noindent\rule{\textwidth}{0.5pt}

\section{Introduction}
\label{sec1}
Dynamic social networks can often be conceptualized as sequences of dyadic interactions between individual entities (e.g., people, institutions, countries, etc.), where the relationships and connections among actors evolve through a series of pairwise exchanges. Within this perspective, social networks are seen as constantly changing systems where the relationships between individuals influence the formation and development of the network. Understanding social networks as a sequence of dyadic interactions provides a nuanced understanding of the dynamic nature of social relationships and the complex interplay between individual interactions and the overall network structure. To this end, a variety of statistical models have been developed to study dynamic social network data \citep{holland_1977_DynamicModelSocial,robins_2001_RandomGraphModels, hanneke_2010_DiscreteTemporalModels,krivitsky_2014_SeparableModelDynamic,leenders_1995_ModelsNetworkDynamics}.

Of particular interest is the relational event model (REM) \citep{butts_2008_RelationalEventFramework,stadtfeld_2014_EventsSocialNetworks,perry_2013_PointProcessModelling}, which provides a framework for analyzing temporal dynamics and patterns within relational event data. Relational event models allow us to estimate and test the relative importance of nodal (e.g., outgoingness or gender) or dyadic attributes (e.g., whether two actors are friends or have common interests) and various types of endogenous interaction mechanisms (e.g., inertia, reciprocity, participation shifts). For example, the endogenous effect ``inertia" quantifies the tendency of actors in the network to keep initiating new events as a function of the volume of past events between the actors. Thus, in the case of a positive inertia effect, if actor A frequently communicated with actor B in the past and only sporadically with actor C, it is more probable that A will communicate with B next than with C. Furthermore, if it is known who are friends with each other (in the form of a dyadic exogenous variable), the REM would allow us to learn how friendships affect social interaction rates while controlling for other effects (such as inertia). By applying the REM on empirical relational event sequences, valuable insights can be gained into social dynamics, such as revealing patterns of reciprocity in friendships and and understanding how communication norms develop over time within social networks.

Traditionally, REMs assume that the predictor variables, such as endogenous statistics (e.g., inertia) or exogenous covariates (e.g., friendships) contribute equally to the interaction rates for all dyads in the network. However, in real life, the drivers of social interaction can vary greatly between dyads. When ignoring this potential source of heterogeneity, the obtained estimates from a fitted REM are simply average effects of predictors (e.g., friendship or inertia) across the entire network. Though useful, these average effects may provide a poor understanding of the actual social interaction mechanisms that are present across all dyads in the network.%  not accurately reflect the relative importance of drivers of social interactions in dynamic networks for all dyads.

To address this, the current paper proposes a dyadic latent class relational event model (DLC-REM) to capture complex social interaction mechanisms caused by dyadic-level variations with the aim to better understand heterogeneous social interaction behavior. The model introduces a latent class approach on a dyadic level which allows two distinct dyads to exhibit distinct patterns of reciprocity, friendship dynamics, or responsiveness to past events. By introducing latent classes, our model goes beyond the traditional homogeneous assumptions of REMs, allowing for the identification of subgroups of dyads that share similar interaction mechanisms. This approach aims to give a more precise understanding of the potential heterogeneity of drivers of social interactions and to yield better predictions in real life social networks.

Alternative approaches in the network literature so far have mainly addressed the unobserved heterogeneity using actor-oriented latent variable and random effects models \citep{mulder2024latent,juozaitiene_2022_NodalHeterogeneityMayb,uzaheta_2023_RandomEffectsDynamica,dubois_2013_StochasticBlockmodelingRelational}. %Typically, simpler variants like using different intercepts to model heterogeneity, common in frailty models \citep{hougaard_1995_FrailtyModelsSurvival} are employed. 
Specifically, latent block-modeling \citep{nowicki_2001_EstimationPredictionStochastic} is a popular method, often used in social networks, to structure the network by partitioning the actors into latent classes. In the case of a static (cross-sectional) network, block-models identify subsets of nodes with similar connectivity. In the case of relational event approaches with block-model structures, the REM parameters are defined by the cluster in which the sender belongs and the cluster in which the receiver belongs \citep{dubois_2013_StochasticBlockmodelingRelational}. Though useful, this approach implies a restricted latent structure on the adjacency matrix, which may result in a poor fit to real life social network data. To improve the fit, one could increase the number of latent classes but this would blow up the number of parameters to be estimated ($TC^2$ for a REM with $T$ exogenous and endogenous effects and $C$ latent blocks), resulting in larger standard errors, and overly complex models which may be difficult to interpret. The proposed DLC-REM, on the other hand, classifies the dyads which behave comparably rather than classifying the individual actors, resulting in $TK$ parameters to be estimated in the case of $K$ dyadic latent classes. The proposed model generalized the stochastic block model by inducing a less restrictive structure on the adjacency matrix. For example, in the case of a friendship covariate, the DLC-REM would be able to identify different levels of friendship relationships (e.g., ``best friends dyads" and ``normal friends dyads") that display different interaction rates. It would be less straightforward to capture such heterogeneity using a stochastic block model \citep[see also][for an interesting discussion of the impact of different latent structures]{hoff2007modeling}. In the current paper, it will be shown how the DLC-REM generalizes the stochastic block model approach using the same number of unknown parameters.

The paper is organized as follows. Section \ref{sec2} provides an overview of the proposed Dyadic Latent Class Relational Event Model (DLC-REM). Section \ref{sec3} delves into the implementation details of the DLC-REM, covering aspects such as model fitting and assessment. Section \ref{sec4} presents numerical simulations to assess the performance of DLC-REM and compared the performance with relational event stochastic block models. In Section \ref{sec5} we analyze interaction dynamics of militarized disputes across countries using the DLC-REM model. The paper concludes with a discussion in Section \ref{sec6}.

\section{Dyadic Latent Class Relational Event Model}
\label{sec2}
\subsection{Model Overview}
%\subsection{Tie-Oriented Relational Event Model}
%Let $y_{md}$ represent a binary variable indicating the occurrence (1) or non-occurrence (0) of an event between dyad $d = (i,j)$ in a network at time point $m$. 

This section presents the DLC-REM by introducing dyadic latent classes within relational event modeling framework of \cite{butts_2008_RelationalEventFramework}.
Formally, we consider $ E = \{ e_1, e_2 \dots \}$, to be a sequence of observed events,  where $e_m = \{ i_m, j_m, t_m\}$ is a tuple of the sender, receiver and time of the $m$-th observed event. The sequence of relational events is modeled using a non-homogeneous multivariate Poisson counting process on the dyads $d=(i,j)$:

\begin{equation}
    \mathbf{N}(t) = (\mathbf{N}_d(t) | d \in \mathcal{R}),
\end{equation}
where $\mathcal{R}$ denotes the riskset of dyads which can be involved in an event. We assume that the riskset does not change over time although this assumption can be relaxed in a straightforward manner. Moreover, throughout this paper relational events are assumed to be directional (i.e., from \textit{i }to \textit{j}) although our approach can naturally be applied to undirected relational events as well. Each element $\mathbf{N}_d(t)$ of $\mathbf{N}(t)$, indicates the cumulative count of events for the dyad $d$, i.e from $i$ to $j$ in the time interval $[0,t)$. Moreover, we define $\Delta \mathbf{N}(t_1, t_2) = \mathbf{N}(t_2) - \mathbf{N}(t_1)$ for $t_2 >t_1$, where the $d$-th element, $\Delta \mathbf{N}(t_1, t_2)$, denotes the count of events that occurred for dyad $d$ in the interval $ [t_{1},t_2)$.

To capture dyadic-level variation in the network, we posit that the dyads in the network belong to a population of $K$ unobserved groups, classes, or clusters. The underlying latent class of a dyad $d$ is denoted by $z_d \in \{1, 2, \ldots, K\}$. Assuming that the dyad belongs to class $k$, i.e $z_d=k$, the interaction rates (intensities) of dyads are specified as a log-linear function of the predictors (also known as statistics) and $T$ class-specific parameters $\boldsymbol{\boldsymbol{\beta}}_{k}$ as follows:
\begin{equation}
    \lambda_{d}(t \ | \ z_d = k, \boldsymbol{\beta}_k, \mathbf{x}_d(E_t)) = \begin{cases} 
     \exp\{  \boldsymbol{\beta}_k^\mathsf{T} \; \mathbf{x}_d(E_t) \} & \  d \in \mathcal{R} \\
    0 & \ d \notin \mathcal{R}
    \end{cases}
    \label{eqn:dlc-rate}
\end{equation}
where $\mathbf{x}_d(E_t)$ is a vector of predictor statistics of dyad $d$ which can contain endogenous statistics (e.g., inertia, reciprocity) that summarize the past activity between actors until  time $t$, exogenous statistics, such as nodal or dyadic attributes, and possible interactions. %The vector $\mathbf{x}_d(E_t)$ represents the statistics for a dyad $d$, given the event history $E_t$ until $t$. The vector of class specific REM parameters for dyadic class $k$ is denoted by $\boldsymbol{\beta}_{k}$.

The likelihood of the relational event sequence comprised of $M$ discrete time intervals ($E_{t_M}$) under the DLC-REM is:
\begin{equation}
    P(E_{t_M} | \boldsymbol{\beta}, \textbf{X}) = \prod \limits_{d \in \mathcal{R}} \; \sum \limits_{k=1} ^{K} \;  P(z_d=k) \; \prod \limits_{m=1}^{M} P(\Delta \mathbf{N}_d(t_{m-1},t_m) \ | \ z_d = k ,\boldsymbol{\beta}_k, \mathbf{x}_d(E_{t_m}))
\end{equation}
% \[
% P(z_d=k) = \pi_k\;\; \;  \text{where} \; \; \pi_k \geq 0 \; \; \text{and} \; \; \sum \limits_{k=1}^{K} \pi_k = 1 ,
% \]
where $P(z_d=k)$ represents the probability of a dyad belonging to class \textit{k}, and $t_0$ denotes the start of the observation period. %In other words, $P(z_d=k)$ is the unknown proportion of the population of dyads in the network that belong to latent class k. 
Further, $P(\Delta \mathbf{N}_d(t_{m-1},t_m) \ | \ z_d = k ,\boldsymbol{\beta}, \mathbf{x}_d(E_{t_m}))$ corresponds to the probability of $\Delta \mathbf{N}_d(t_{m-1},t_m)$ events in the time interval $[t_{m-1},t_m)$ for dyad $d$ given class membership $z_d=k$. The probability of observing an event (i.e $\Delta \mathbf{N}_d(t_{m-1},t_m) = 1$)  in the time interval for $d$ given $z_d$ can be specified under a Poisson distribution as:

\begin{flalign}
&P(\Delta \mathbf{N}_d(t_{m-1},t_m) = 1 \mid z_d = k, \boldsymbol{\beta}_k, \mathbf{x}_d(E_{t_m})) = &\nonumber \\
& \hspace{4cm}\quad \frac{\lambda_{d}(t_m, \mathbf{x}_d(E_{t_m}) \mid z_d = k, \boldsymbol{\beta})^n \ \exp\{-\lambda_{d}(t_m, \mathbf{x}_d(E_{t_m}) \mid z_d = k, \boldsymbol{\beta}_k)\}}{n!}.
\end{flalign}

Often, relational event datasets may contain events that (i) occur concurrently or (ii) the resolution of the available timing of events doesn't allow one to determine the exact order of several events within a period. For instance, in social environments, it is common for several conversations to happen at once, or for multiple simultaneous reactions to occur in response to a single event, complicating the analysis of individual interactions. In DLC-REM, the change in the network between intervals is denoted by the variable $\Delta \mathbf{N} (t_{m-1},t_{m})$. Consequently, multiple events may be observed involving various dyads within the same time interval. This way, no arbitrary order is assumed for the events that occurred in the same interval. Furthermore, it is postulated that only the events that occurred preceding the start of this interval are assumed to affect the occurrence of an event within the current interval. In practice, this means that instead of updating the statistics matrix 
$\mathbf{x}_d(E_m)$ after observing each individual event, as in traditional Relational Event Models, we update the statistics matrix simultaneously for all the events observed within the interval $[t_{m-1},t_m)$.

\subsection{Concomitant Model}
Latent class models can be extended to include a concomitant model \citep{dayton_1988_ConcomitantVariableLatentClassModels}.
The concomitant model specifies the probability of belonging to a specific latent class as a function of the (chosen) concomitant variables, $\textbf{w}$, and corresponding coefficients $\boldsymbol\gamma$. For the DLC-REM, a concomitant model can be added to model the class memberships of all active dyads in the network. The class membership probabilities of the concomitant model are parameterized using a multinomial logistic regression model:
\begin{equation}
    P(z_d=k | \textbf{w}_d, \boldsymbol\gamma_k) = \frac{\exp(\textbf{w}'_d \ \boldsymbol\gamma_k)}{\sum \limits_{z=1}^{K} \exp(\textbf{w}'_d \ \boldsymbol\gamma_z) }.
\end{equation}
Therefore in addition to the $ \{ \boldsymbol{\beta}_1,\dots \boldsymbol{\beta}_k \}$ we would also have to estimate $ \{ \gamma_1,\dots \gamma_{k} \}$

The concomitant models allows social network researchers (i) to establish and understand the relationships between the concomitant variables and the grouping of observations into latent classes, which can help in identifying which characteristics are predictive of class membership, and (ii) to investigate the expectations or typical values of the concomitant variables within each class, thereby profiling or characterizing each latent class based on observable variables \citep{wedel_2002_ConcomitantVariablesFinite}. The concomitant variables for the DLC-REM may be a subset of or be completely disjoint from the statistics utilized to model the relational events. 

The concomitant model can also be fitted in a separate step, as suggested by \cite{vermunt_2017_LatentClassModeling}. In this approach, the class memberships are assumed to be known, which simplifies the estimation process. However, this separate estimation may result in an overestimation of the certainty of class membership, as the potential uncertainty or variability in class assignment is not fully accounted for. This is one of the reasons why fitting the model in one step, where the class memberships and concomitant model are estimated simultaneously, is often preferred. We will revisit this issue in the Discussion.

\subsection{Relation to Stochastic Block-Models}
Blockmodels have long served social scientists for dissecting social and relational structures. The relational event stochastic blockmodel (which we refer to as SB-REM) introduced by \cite{dubois_2013_StochasticBlockmodelingRelational} provides a blockmodelling approach for relational event models. In the SB-REM, the actors are assumed to be allocated in one out of $C$ possible latent classes. The parameter vector $\boldsymbol{\theta}_{c_1,c_2}$ of length $T$ quantifies the relative importance of the network drivers of social interactions of a sending actor in block $c_1$ towards a receiving actor in block $c_2$. These parameters have a comparable interpretation as the parameter vector $\boldsymbol{\beta}_k$ in the DLC-REM which quantifies the relative important of the social interaction process for dyads in latent dyadic class $k$.

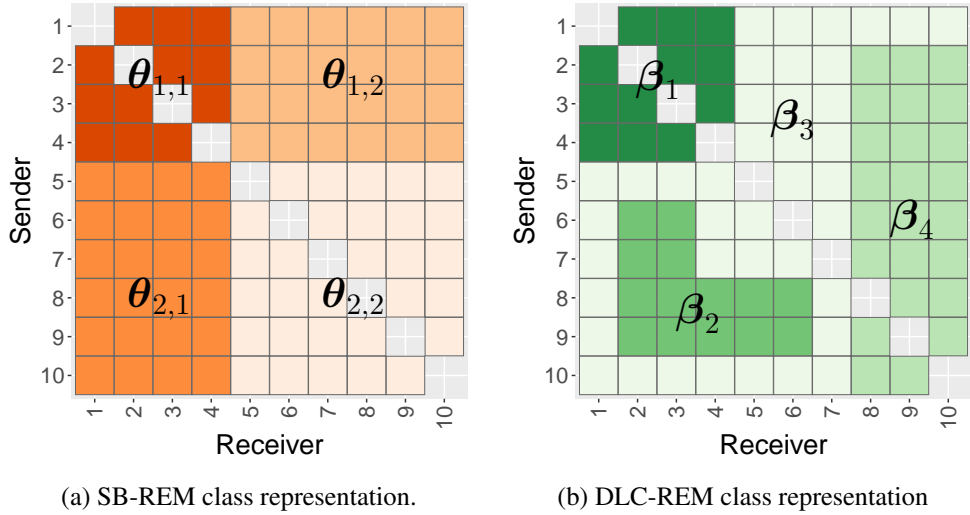
\begin{figure}[t]
    \centering
      \begin{minipage}{0.4\textwidth}
    \centering
    \begin{tikzpicture}
    % Insert the image
    \node[anchor=south west, inner sep=0] (image) at (0,0) {\includegraphics[width=\textwidth]{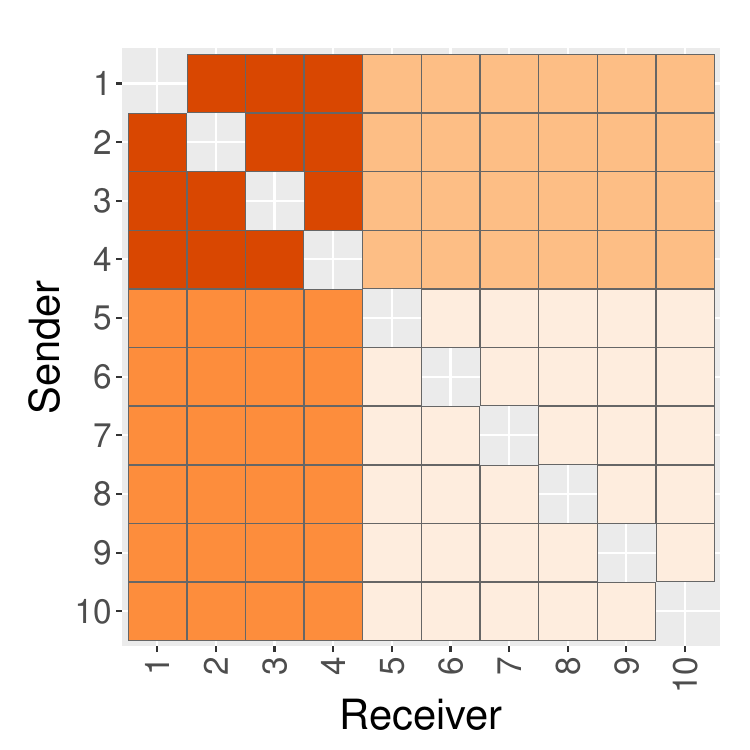}};
    \begin{scope}[x={(image.south east)}, y={(image.north west)}]
        % Add text annotations at various positions with increased font size
        \node[font=\fontsize{16}{12}\selectfont] at (0.34, 0.78) {$\boldsymbol{\theta}_{1,1}$};  
        \node[font=\fontsize{16}{12}\selectfont] at (0.34, 0.34) {$\boldsymbol{\theta}_{2,1}$};  
        \node[font=\fontsize{16}{12}\selectfont] at (0.73, 0.78) {$\boldsymbol{\theta}_{1,2}$};  
        \node[font=\fontsize{16}{12}\selectfont] at (0.73, 0.34) {$\boldsymbol{\theta}_{2,2}$};  
    \end{scope}
\end{tikzpicture}
    \subcaption{SB-REM class representation.}
    \label{plot:relation_sbm}
  \end{minipage}
  \begin{minipage}{0.4\textwidth}
    \centering
    \begin{tikzpicture}
    % Insert the image
    \node[anchor=south west, inner sep=0] (image) at (0,0) {\includegraphics[width=\textwidth]{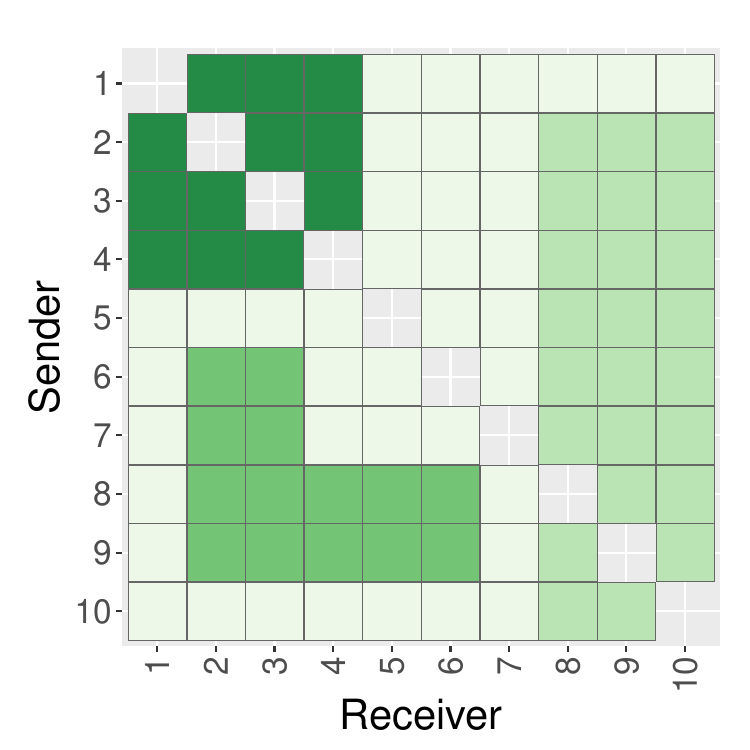}};
    \begin{scope}[x={(image.south east)}, y={(image.north west)}]
        % Add text annotations at various positions with increased font size
        \node[font=\fontsize{16}{12}\selectfont] at (0.33, 0.78) {$\boldsymbol{\beta}_1$};  % Very large font for $\beta_1$
        \node[font=\fontsize{16}{12}\selectfont] at (0.41, 0.31) {$\boldsymbol{\beta}_2$};  % Very large font for $\beta_2$
        \node[font=\fontsize{16}{12}\selectfont] at (0.6, 0.7) {$\boldsymbol{\beta}_3$};  % Very large font for $\beta_3$
        \node[font=\fontsize{16}{12}\selectfont] at (0.84, 0.5) {$\boldsymbol{\beta}_4$};  % Very large font for $\beta_4$
    \end{scope}
\end{tikzpicture}
    \subcaption{DLC-REM class representation}
    \label{plot:relation_lc}
  \end{minipage}

  \caption{Depiction of SB-REM and SLC-REM latent classes}
  \label{plot:relation}
  \end{figure}

It can be argued that the SB-REM is a special case of the DLC-REM because the implied adjacency matrix of a SB-REM can also be created a DLC-REM with the same number of parameters but the implied adjacency matrix of a DLC-REM generally cannot be created by a SB-REM with the same number of parameters. To see this, we consider a SB-REM with two blocks for a network of 10 actors where the first 4 actors belong to latent block 1 and the last 6 actors belong to block 2  (as depicted in the adjacency matrix in Figure \ref{plot:relation}(\subref{plot:relation_sbm})), there are four sets of parameters, $\theta_{1,1}$, $\boldsymbol{\theta}_{1,2}$, $\boldsymbol{\theta}_{2,1}$, and $\boldsymbol{\theta}_{2,2}$, where (for instance) $\boldsymbol{\theta}_{2,1}$ denotes the REM parameters for interactions of sender actors in block 2 (either 5, 6, $\ldots$, or 10) towards receivers actors belonging to block 1 (either 1, 2, 3, or 4). It is straightforward to recreate this latent structure using a DLC-REM using four dyadic classes with four sets of parameters $\boldsymbol{\beta}_1, \boldsymbol{\beta}_2, \boldsymbol{\beta}_3$, and $ \boldsymbol{\beta}_4$, where (for instance) $\boldsymbol\beta_3$ contains the parameters for the dyads of which the sender can be actor 5 to 10, and the receiver can be actor 1 to 4. Alternatively, we consider a DLC-REM with four latent dyadic classes and the adjacency configuration in Figure(\ref{plot:relation}(\subref{plot:relation_lc}). This DLC-REM on the other hand cannot be captured by a SB-REM with 2 latent blocks.

The DLC-REM also generalizes the SB-REM due to its greater flexibility in the number of REM parameters. The number of parameter vectors in the SB-REM is equal to $C^2$, where C is the total number of blocks for the actors. Consequently, the total number of parameter vectors will be equal to $4, 9, 16, \dots$ in the case of $C = 2, 3, 4, \dots$ blocks for the actors, respectively. In contrast, the DLC-REM allows the total number of parameter vectors to be any integer $K = 1, 2, 3, 4, 5, \dots$.

In sum, the DLC-REM can be viewed as a generalization of the SB-REM because it is more flexible regarding (i) the network configurations of latent class parameters and (ii) the number of parameter vectors of the latent classes that it may contain. See also \cite{dubois_2010_ModelingRelationalEvents} for a related discussion of their multinomial mixture model for relational event frequencies. Section \ref{sim} presents several numerical simulations to compare the DLC-REM with the SB-REM.

\section{Technical Implementation of the DLC-REM}
\label{sec3}
In the previous section, we translated the dyadic latent class approach for relational event data into a mathematical framework. In this section, we further discuss the implementation details of the model and the estimation procedure utilized to estimate the model.

\subsection{Model Fitting}
The DLC-REM can be estimated using the EM algorithm \citep{dempster_1977_MaximumLikelihoodIncomplete} given a fixed number of classes K. The ML estimates of the joint rate and concomitant model $\{ \mathcal{B}, \Gamma \}$ are given by maximizing the following log-likelihood:

\begin{equation}
\log P(E_{t_M} | \boldsymbol{\beta}, \boldsymbol{\gamma}, \textbf{X}) =  \sum \limits_{d \in \mathcal{R}} log \left\{   \sum \limits_{k=1}^{K} P(z_d=k | \textbf{w}_d, \boldsymbol{\gamma}_k) \; \prod \limits_{m=1}^{M} P(\Delta \mathbf{N}_d(t_{m-1},t_m) \ | \ z_d = k ,\boldsymbol{\beta}, \mathbf{x}_d(E_{t_m})) \right\}
\end{equation}

\textbf{E-Step:} \newline
Posterior class probabilities that dyad $d$ belongs to class $k$ are estimated given the current parameter estimates of the i-th iteration $\hat{\boldsymbol{\beta}}^{(i)}$ and $\hat{\boldsymbol{\gamma}}^{(i)}$: 

% \begin{align}
% \label{eqn:post-concom}
%     \hat{p_{dk}} =  P(z_d =k,\hat{\boldsymbol{\beta}}, \hat{\gamma} |\mathbf{x}_d(E_{t_m})) = \frac{P(z_d=k | \textbf{w}_d, \hat{\gamma_k}) \prod \limits_{m=1}^{M} P(\Delta \mathbf{N}_d(t_{m-1},t_m) | z_d=k , \hat{\boldsymbol{\beta}}_{k}, \mathbf{x}_d(E_{t_m})) ) }{  \sum \limits_{h=1}^{K} P(z_d=h | \textbf{w}_d, \hat{\gamma_h}) \prod \limits_{m=1}^{M} P(\Delta \mathbf{N}_d(t_{m-1},t_m) | z_d = h , \hat{\boldsymbol{\beta}}_h, \mathbf{x}_d(E_{t_m})) )}
% \end{align}

\begin{align}
\label{eqn:post-concom}
   \hat{p}_{dk} = & \frac{P(z_d=k | \textbf{w}_d, \hat{\boldsymbol{\gamma}}_k^{(i)}) \prod \limits_{m=1}^{M} P(\Delta \mathbf{N}_d(t_{m-1},t_m) | z_d=k , \hat{\boldsymbol{\beta}}_{k}^{(i)}, \mathbf{x}_d(E_{t_m})) }{ \sum \limits_{h=1}^{K} P(z_d=h | \textbf{w}_d, \hat{\boldsymbol{\gamma}_h}^{(i)}) \prod \limits_{m=1}^{M} P(\Delta \mathbf{N}_d(t_{m-1},t_m) | z_d = h , \hat{\boldsymbol{\beta}}_h^{(i)}, \mathbf{x}_d(E_{t_m})) }
\end{align}

% \begin{align}
% \label{eqn:post-concom}
%    \hat{p}_{dk} = & \ P(z_d =k,\hat{\boldsymbol{\beta}}, \hat{\boldsymbol{\gamma}} |\mathbf{x}_d(E_{t_m})) \nonumber \\
%     = & \frac{P(z_d=k | \textbf{w}_d, \hat{\boldsymbol{\gamma}_k}) \prod \limits_{m=1}^{M} P(\Delta \mathbf{N}_d(t_{m-1},t_m) | z_d=k , \hat{\boldsymbol{\beta}}_{k}, \mathbf{x}_d(E_{t_m})) }{ \sum \limits_{h=1}^{K} P(z_d=h | \textbf{w}_d, \hat{\boldsymbol{\gamma}_h}) \prod \limits_{m=1}^{M} P(\Delta \mathbf{N}_d(t_{m-1},t_m) | z_d = h , \hat{\boldsymbol{\beta}}_h, \mathbf{x}_d(E_{t_m})) }
% \end{align}

\textbf{M-Step:} \newline
The estimation of the $\boldsymbol{\beta}$ and $\boldsymbol{\gamma}$ parameters can be done separately because the conditional expectation of the log likelihood $Q$ in the i-th iteration is given by the following equation which can be decomposed into two distinct parts, each involving only one of the parameter sets:

% \[
% E(log L) = \sum \limits_{d \in \mathcal{R}} \sum \limits_{k=1}^{K} \hat{p}_{dk}\;  log \left\{ P(z_d=k | \textbf{w}_d, \gamma_k) \right\} \; \; + \; \; \sum \limits_{d \in \mathcal{R}} \sum \limits_{k=1}^{K} \hat{p}_{dk}\;  log \left\{ \prod \limits_{m=1}^{M} P(\Delta \mathbf{N}_d(t_{m-1},t_m) | z_d = k , \boldsymbol{\beta}_k, \mathbf{x}_d(E_{t_m})) \right\}
% \]
\begin{align*}
Q = & \sum \limits_{d \in \mathcal{R}} \sum \limits_{k=1}^{K} \hat{p}_{dk}\;  \log \left\{ P(z_d=k | \textbf{w}_d, \boldsymbol{\gamma}_k) \right\} \\
& + \sum \limits_{d \in \mathcal{R}} \sum \limits_{k=1}^{K} \hat{p}_{dk}\;  \log \left\{ \prod \limits_{m=1}^{M} P(\Delta \mathbf{N}_d(t_{m-1},t_m) | z_d = k , \boldsymbol{\beta}_k, \mathbf{x}_d(E_{t_m})) \right\}
\end{align*}

Therefore, using the estimates of the posterior probabilities $\hat{p}_{dk}$ as weights, one can obtain the new estimates of the parameters by maximizing the following two equations: 
\begin{equation}
    \max \limits_{\boldsymbol{\beta}} \sum \limits_{d \in \mathcal{R}} 
 \sum \limits_{k=1}^{K}  \hat{p}_{dk} \  log  \prod \limits_{m=1}^{M} P(\Delta \mathbf{N}_d(t_{m-1},t_m) | z_d = k , \boldsymbol{\beta}_k, \mathbf{x}_d(E_{t_m}))
 \label{eqn:m-beta}
\end{equation}

\begin{equation}
    \max \limits_{\boldsymbol{\gamma}} \sum \limits_{d \in \mathcal{R}} 
 \sum \limits_{k=1}^{K}   \hat{p}_{dk} \  log(P(z_d=k | \textbf{w}_d, \boldsymbol{\gamma}_k))
 \label{eqn:m-gamma}
\end{equation}

Equation \ref{eqn:m-beta} is maximized using maximum likelihood estimates of GLMs and Equation \ref{eqn:m-gamma}  is maximized using the Maximum likelihood estimation of multinomial logit models. 

Like traditional latent class models, the likelihood function of the DLC-REM can exhibit multiple modes. Consequently, executing the EM algorithm just once might lead to sub-optimal solutions. To assess the identifiability of the model when employing the EM algorithm, one effective method is to perform the estimation using multiple initializations. If these varied initializations lead to identical log-likelihood values but yield divergent parameter estimates, it indicates that the model suffers from issues of non-identifiability \citep{vermunt_1996_LoglinearEventHistory}.

\subsection{Model assessment}
Identifying the optimal number of latent classes that best represent the underlying structure of dyadic interactions, is one of the most critical tasks in the analysis of latent class models. To identify the optimal number of classes in a DLC-REM model, various criteria can be employed, such as the Akaike Information Criterion (AIC) or Bayesian Information Criterion (BIC). These criteria assess the trade-off between model fit and complexity, to determine the most suitable number of latent classes. AIC tends to be quite liberal in selecting the number of classes, often favoring models with a higher number of latent classes. On the other hand, the BIC is more conservative, typically selecting a more parsimonious model with fewer classes. This conservatism of the BIC can be particularly useful in preventing over-fitting \citep{nylund_2007_DecidingNumberClasses}.

The predictive performance of relational event models can also be used to assess model fit to determine the optimal number of classes. To do this, we calculate the estimated rate for each dyad according to the fitted model and rank the dyads in decreasing order of event rate. We assign rank 1 to the dyad with the highest estimated rate, rank 2 to the dyad with the second largest estimated rate, and so forth. At each time point we observe the rank of the dyad that actually occurred in the observed event sequence. If the rank of the ``realized'' i.e the observed event is greater than a threshold percentile, then that event is considered correctly predicted by the model. The proportion of correctly predicted events (also referred to as \textit{recall}) can be used as a predictive performance \citep{dubois_2013_HierarchicalModelsRelational,gravel_2023_RivalriesReputationRetaliation} measure for comparing the goodness of fit of relational event models.

The predictive performance measure is particularly useful and can be even more insightful than the AIC or BIC because it provides an absolute measure of model fit rather than a relative one. Unlike AIC and BIC, which balance model fit with model complexity, this measure directly evaluates how well the model predicts actual events in the data. Its scale has a more intuitive interpretation—specifically, the proportion of correctly predicted events. 

\begin{figure}[htbp]
  \begin{minipage}{0.3\textwidth}
    \centering
    \includegraphics[width=\linewidth]{gen_inertia_r_1_1.pdf}
    \subcaption{Data Generating Structure}
    \label{plot:sim_gen}
  \end{minipage}
  \hfill
  \begin{minipage}{0.3\textwidth}
    \centering
    \includegraphics[width=\linewidth]{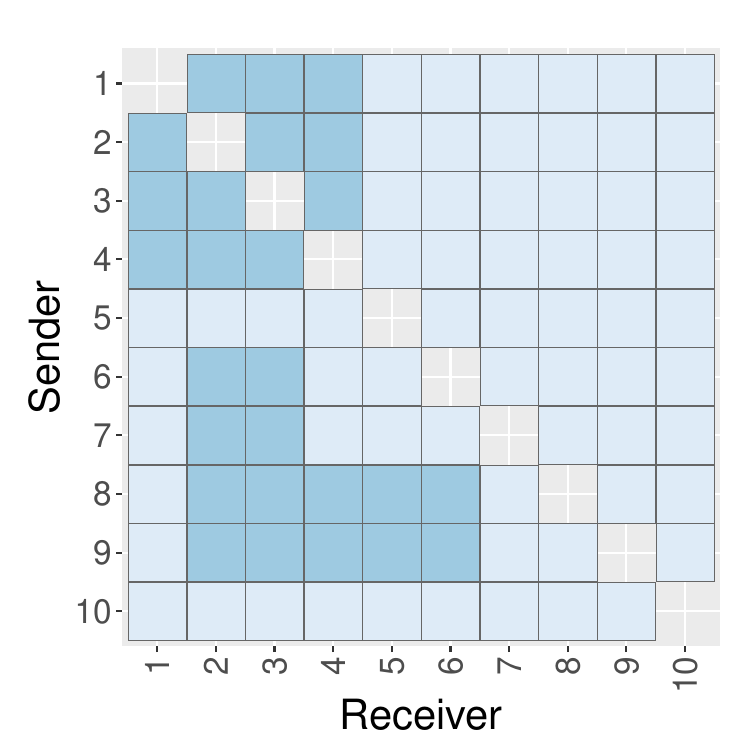}
    \subcaption{Fitted dyadic LC, K=2}
    \label{plot:sim_lc_2}
  \end{minipage}
  \hfill
  \begin{minipage}{0.3\textwidth}
    \centering
    \includegraphics[width=\linewidth]{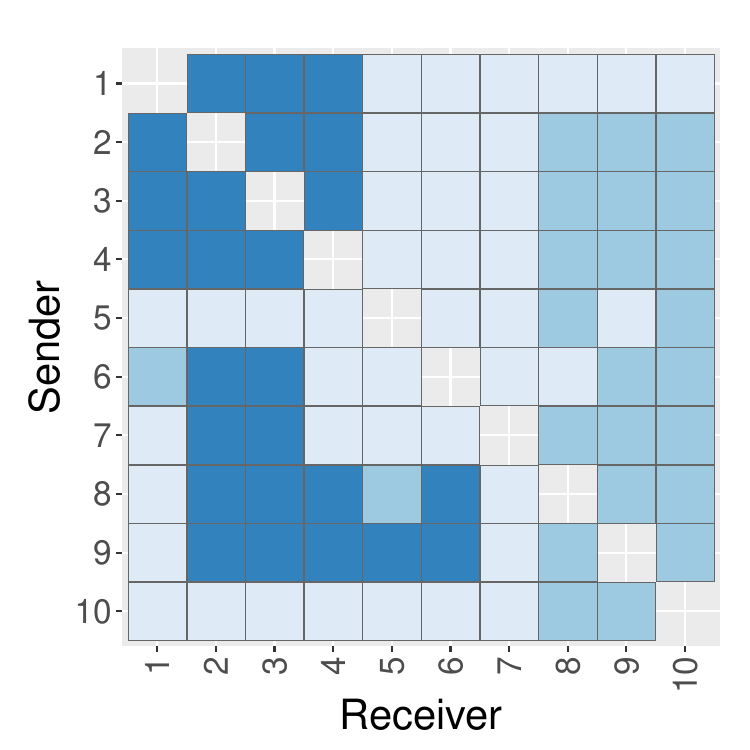}
    \subcaption{Fitted dyadic LC, K=3}
    \label{plot:sim_lc_3}
  \end{minipage}
  \vfill
  \begin{minipage}{0.3\textwidth}
    \centering
    \includegraphics[width=\linewidth]{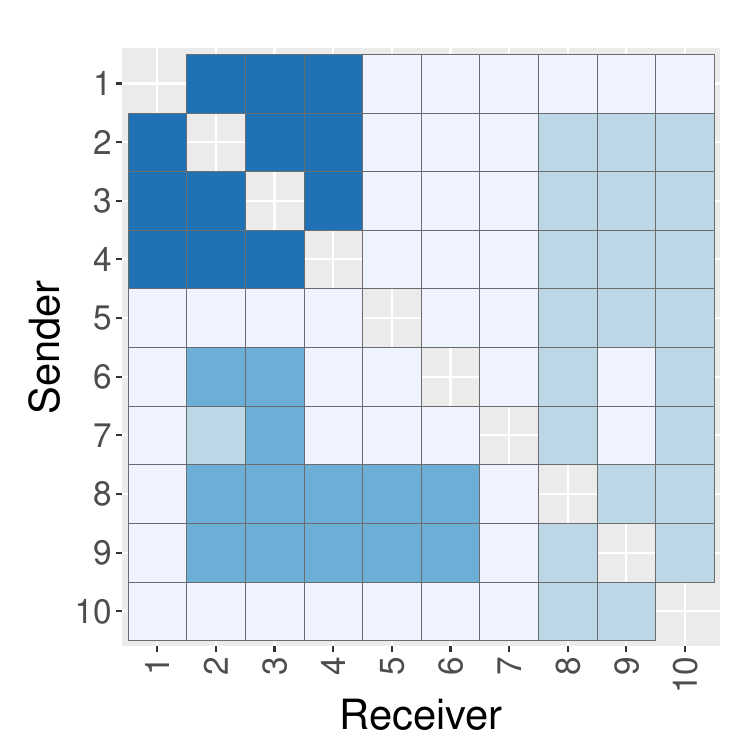}
    \subcaption{Fitted dyadic LC, K=4}
    \label{plot:sim_lc_4}
  \end{minipage}
  \hfill
  \begin{minipage}{0.3\textwidth}
    \centering
    \includegraphics[width=\linewidth]{sbm_fit_inertia_K=2_r_1_1.pdf}
    \subcaption{Fitted BM, K=2}
    \label{plot:sim_sbm_2}
  \end{minipage}
  \hfill
  \begin{minipage}{0.3\textwidth}
    \centering
    \includegraphics[width=\linewidth]{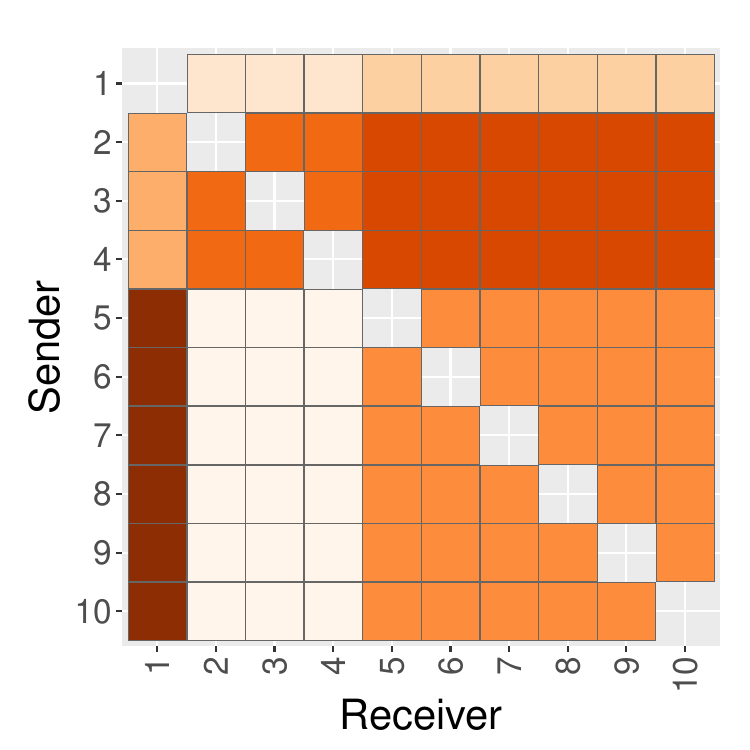}
    \subcaption{Fitted BM, K=3}
    \label{plot:sim_sbm_3}
  \end{minipage}
  \hfill
  \begin{minipage}{0.3\textwidth}
    \centering
    \includegraphics[width=\linewidth]{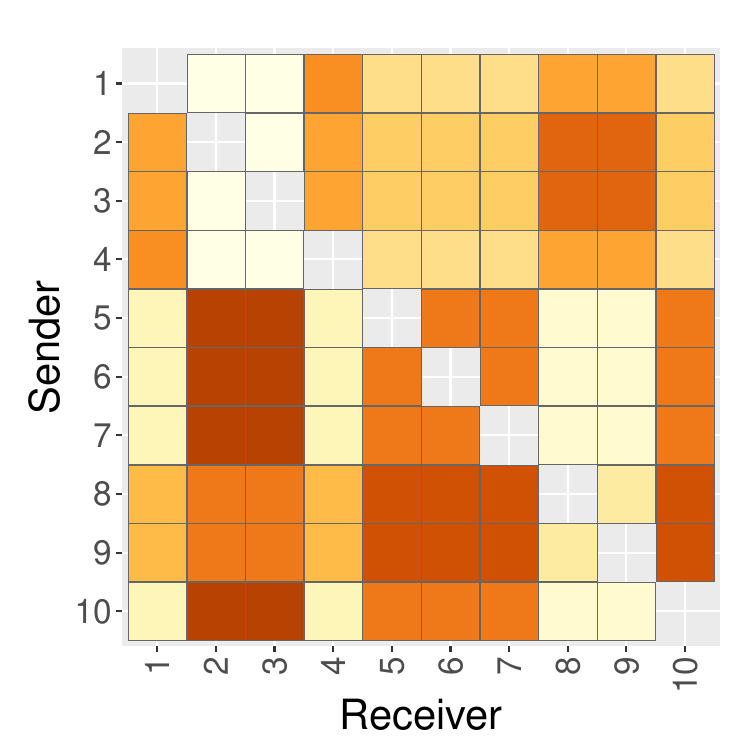}
    \subcaption{Fitted BM, K=4}
    \label{plot:sim_sbm_4}
  \end{minipage}
\hfill
  \begin{minipage}{0.65\textwidth}
    \centering
    \begin{tabular}{ccc}
      \toprule
      model & recall & \#param vectors\\
\midrule
REM & 0.66 & 1 \\
DLC-REM K=2 & 0.72 & 2\\
DLC-REM K=3 & 0.73 & 3 \\
DLC-REM K=4 & 0.74 & 4 \\
SB-REM C=2 & 0.68 & 4\\
SB-REM C=3 & 0.71 & 9 \\
SB-REM C=4 & 0.72 & 16 \\
      \bottomrule
    \end{tabular}
    \subcaption{Predictive performance}
    \label{tab:bm_lc}
  \end{minipage}
  \caption{(a) The data generating structure used in simulations (b,c,d) The structure of a fitted DLC-REM  with $K=2,3,4$ respectively. (e,f,g) Structure derived from fitted SB-REMs with 2,3, and 4 classes respectively. Table (h) depicts the recall (under a 95th percentile threshold) and the number of parameter vectors estimated per statistic for each model.}
    \label{plot:sim}
\end{figure}

\subsection{Implementation in R}
\label{sec:flexmix}
Software for fitting latent class models are widely available, that we can make use of when fitting the DLC-REM \citep{fraley2009mclust,grun_2008_FlexMixVersionFinite,benaglia_2009_MixtoolsPackageAnalyzing,9b3146187288472b86c688616ca726fe}.

 The \texttt{prepare\_reh\_glm} function (see Appendix \ref{appendix:b} for the R code) is designed to pre-process relational event data for Generalized Linear Model (GLM) analysis. This function facilitates the construction of a data stack that captures the occurrence or non-occurrence of events between dyads at each time point where an event is observed. This pre-processing step converts traditional relational event data, often represented as separate edgelists and statistical arrays, into a format suitable for fitting finite mixture models within the GLM framework. Specifically, it employs a Poisson family to model the count data typical of relational event models, with the logarithm of interevent times serving as an offset in the Poisson regression \citep[e.g.,][]{holford1980analysis,laird1981covariance,vieira2024fast}.

After this pre-processing step, the latent class model can then be fitted using the \texttt{flexmix} package in R \citep{grun_2008_FlexMixVersionFinite}. The following code snippet demonstrates how the rate model and the concomitant model can be specified for a DLC-REM:
\begin{lstlisting}[language=R,numbers=none]
 lc_fit <- flexmix::flexmix(obs ~ (1 + inertia + reciprocity + ... | tie), 
    data = reh_glm$stat_glm, 
    k = 4,
    model = FLXMRglm(family = "poisson", offset = reh_glm$log_intevent),
    concomitant = FLXPmultinom(~ contiguity_c + alliance_c + ...))

\end{lstlisting}

In this example, the model formula specifies the inclusion of various predictors (e.g., inertia, reciprocity) within the finite mixture model. Additionally, the concomitant model can also be specified, which includes covariates such as \texttt{contiguity\_c} and \texttt{alliance\_c}.

It is also possible to impose more restrictive structures within the latent class model, such as ensuring that both directions of interaction between the same pair of actors belong to the same latent class. This is useful when interaction dynamics are assumed to be symmetric on a dyadic level. In the context of implementation, this could mean ensuring that the dyad id represented by the \texttt{tie} variable from actor A to actor B is identical to the tie from actor B to actor A. 

\section{Numerical Simulations as a Proof of Concept}
\label{sim}
\label{sec4}
In this section we present numerical simulations to assess the performance of DLC-REMs. In particular, we assess whether the model is able to recover the true number of classes. We also compare the performance of the dyadic latent class models to the relational event stochastic blockmodels (SB-REM)\citep{dubois_2013_StochasticBlockmodelingRelational}. 

Using the R package \texttt{remulate} \citep{lakdawala_2024_SimulatingRelationalEvent}, we simulate relational event sequences for a relational event model with a heterogeneous latent structure. The parameters of the REM simulations vary across the dyads and can be visualized using a structure depicted in Figure \ref{plot:sim}(\subref{plot:sim_gen}). We generate 100 relational event sequences (with 2000 events each) from this model such that coefficient for network effects of intercept, inertia and reciprocity vary across the four classes $\boldsymbol{\beta}_{intercept} =  \{-11,-2,-5,-3\}$, $\boldsymbol{\beta}_{inertia} =  \{-0.2,0.1,0.6,0.3\}$, and $\boldsymbol{\beta}_{reciprocity} =  \{-0.3,0.05,0.1,0.2\}$. In addition, we also include effects for participating shifts $\boldsymbol{\beta}_{ps-abba} \boldsymbol{\beta}_{ps-abby}, \ \boldsymbol{\beta}_{ps-abay}$ (see R code in Appendix \ref{appendix:a} for details).
 
We then fit a DLC-REM and a SB-REM\footnote{The SB-REM is fitted using an extension of the R implementation provided by \citep{dubois_2013_StochasticBlockmodelingRelational}} on these simulated networks. Figures \ref{plot:sim}(\subref{plot:sim_lc_2},\subref{plot:sim_lc_3}) shows the fitted structures on one of the simulated sequences 2-class and a 3-class DLC-REM for the inertia effect. Although these simpler models partly capture the latent structure, part of the data generating structure is lost when choosing fewer classes than the true model. Figure \ref{plot:sim}(\subref{plot:sim_lc_4}) shows the structure resulting from a fitted DLC-REM for $K=4$ which corresponds to the true value of distinct groups in the data. The results indicate that the dyadic latent class model is able to capture the original structure very well, with only a few mis-classified dyads.

Figures \ref{plot:sim}(\subref{plot:sim_sbm_2},\subref{plot:sim_sbm_3},\subref{plot:sim_sbm_4}) show the structure derived from fitted block models with 2, 3, and 4 blocks respectively. However, it is clear from Figure \ref{plot:sim}(\subref{plot:sim_sbm_2}) that such a blockmodel does not capture the data-generating structure well. The SB-REM with three blocks also cannot capture the underlying data generating structure adequately whereas the SB-REM with four classes can model the data generating structure but with too many partitions (16 parameters) which results in a model that is not parsimonious. Table (\subref{tab:bm_lc}) under Figure \ref{plot:sim} displays the average recall of various models across the 100 simulated sequences under a $95^\text{th}$ percentile threshold. The standard relational event model, that is the same as a dyadic latent class model with one class, is also included as a baseline. Table (\subref{tab:bm_lc}) under Figure \ref{plot:sim} also reports the number of parameter vectors that are estimated per statistic included in the model.  The results indicate that the DLC-REM has superior predictive performance compared to the SB-REM and standard REM while also maintaining a more parsimonious model. 

These simulations illustrated the increased flexibility of the proposed dyadic latent class approach, in particularly in the relational event context and in contrast to traditional block-model approaches. 

\section{Analyzing interaction dynamics from militarized disputes data}
\label{sec5}
We will demonstrate the utility of our dyadic latent class model to study complex interaction dynamics between countries in international militarized conflict. It is well-known that pairs of states or dyads have unique relations and mechanisms in the context of dispute dynamics. Few analyses of militarized interstate dispute data have investigated the dynamics of the disputes by modelling the occurrence of the disputes themselves. To our knowledge, these datasets have not yet been analyzed using relational event models. The DLC-REM has the ability to identify different interaction styles in the context of international conflicts which will likely differ between pairs of states.  Accounting for this heterogeneity across dyads is imperative to truly understand complex interaction processes in dynamic networks. The approach of our proposed dyadic latent class relational event model is ideal for this research setting as it combines the powerful interpretation of the relational event model while accounting for the unobserved differences of interaction styles across pairs of states. In this section, we first describe the data, then we discuss the specification of the model, the choice of the number of latent dyadic classes, and the interpretation of the fitted model. As a reference analysis, the fitted DLC-REMs are compared with the fit of a standard REM without any latent variables. To our knowledge no software is currently available for fitting SB-REMs with the endogenous statistics that are included in the model, and therefore, SB-REMs are not considered in this section.

\subsection{Description of the datasets}
In this study we investigate the dynamics of militarized interstate disputes. A militarized interstate dispute (MID) is defined as ``a set of incidents involving the deliberate, overt, government-sanctioned, and government-directed threat, display, or use of force between two or more states.'' \citep{maoz_2019_DyadicMilitarizedInterstate}. The Correlates Of War project (COW) is a dataset repository that is a comprehensive and widely used resource in the field of international relations and political science \citep{singer_1972_CorrelatesWarProject}. The COW dataset offers a collection of data related to international disputes, cooperation, geopolitics and other measures of international behavior among 190 countries from 1812-2014. We focus on the dyadic Militarized Interstate Disputes dataset (version 4.0) and we pooled data from five other datasets in this study: the National Material Capabilities (version 6.0), Formal Alliances (version 4.1), Correlates of War Project’s Trade Data Set (version 4.0), Correlates of War Direct Contiguity data (version 3.20) and Polity5: Political Regime Characteristics
and Transitions, 1800-2018 to specify the network effects in our model.

The National Material Capabilities (NMC) dataset \citep{singerj.david_1972_CapabilityDistributionUncertainty} contains information on various indicators of material capabilities of nations in the COW project.% covering an expansive time period 1816-2012. % The indicators are categorized on three dimensions: demographic, industrial and military. 
 The dataset also provides a consolidated Composite Index of National Capability (CINC) score \citep{singer_1972_CapabilityDistributionUncertainty} for each state. This score aggregates six individual measures into a single value per year per state. The CINC score quantifies national capabilities using 6 indicators of national material capabilities (military personnel count, military expenditure, iron and steel production, energy consumption, total population, and total urban population), each of which are essential for assessing a country's potential to initiate or sustain militarized disputes. Another dataset that is used in this application is the Trade dataset \citep{barbieri_2009_TradingDataEvaluating}. This dataset provides various measures of trade between countries. We focus on the flow of imports from country A to B in million US dollars. The Direct Contiguity dataset \citep{stinnett_2002_CorrelatesWarCow} contains information on the direct contiguity relationship between states in the COW project. The direct contiguity is divided into categories based on land and water contiguity. The Formal Interstate Alliance dataset \citep{giblerdouglasm._2009_InternationalMilitaryAlliances16482008} contains information on formal alliances between states over an extensive period of 1816-2012. A formal alliance is a written agreement that identifies at least the members and minimal obligations of signatories \citep{singer_1966_FormalAlliances1815,gibler_2004_MeasuringAlliancesCorrelates}.  The formal alliances are coded into 3 types:- 'I' defense pact, 'II' neutrality or non-aggression pact, or 'III' entente. Generally, Type I alliances impose a higher level of obligation on the allied states  than the Type II or Type III alliances. In our analysis, we focus on Type I defense alliances that require member states to intervene with a military defense on the side of any treaty partners if attacked by a foreign state. We also utilize the Polity5 dataset \citep{marshall_2020_CenterSystemicPeace} that provides a democracy score (scale 0-10) and an autocracy score (0-10). These external sources of information are used to construct exogenous statistics to model interaction rates of military disputes between states.
%We also utilize the World Religion Dataset \citep{maoz_2013_WorldReligionDataset}. The data provides  percentages of a state’s population that practice a given religion from 1945-2010. We use this dataset to determine the predominant religion in each state. For periods preceding 1945, we extrapolate the majority religion back to 1914.

\renewcommand{\arraystretch}{1.8}
\begin{table}[htbp]
    \centering
    \fontsize{10pt}{10pt}\selectfont
    \begin{tabular}{>{\raggedright\arraybackslash}p{0.35\textwidth} >{\arraybackslash}p{0.65\textwidth}}
        \toprule
        \textnormal{Network statistic} & Interpretation: \newline A positive effect of this statistic implies that \dots \\ 
            \midrule          
            $^{*}$ Inertia &  dyads involved in an MID in the past have higher rate of being involved in an MID. \\
            $^{*}$ Reciprocity & initiators have higher rates of initiating MIDs to targets who have targeted them in MIDs in the past.\\
            $^{*}$ Recent target persistence (rrankSend) & initiators that have recently initiated an MID towards the target, have a higher rate of initiating an MID towards the target.\\
            $^{*}$ Recent retaliation (rrankReceive) & targets that have recently been targeted by an initiator, have a higher rate of initiating an MID towards the initiator.\\
            $^{*}$ Displaced conflicts (ps ab-by) & targets of an MID that are attacked by a specific initiator are more likely to immediately attack a third actor in subsequent MIDs. \\
            $^{*}$ Violent spree (ps ab-ay) & initiators of an MID that attacked a specific target are more likely to immediately attack a third actor in subsequent MIDs. \\
            %$^{*}$ Out Degree Initiator &   initiators that initiated a higher volume of MIDs in the past, have higher rates of initiating MIDs.\\
            %$^{*}$ In Degree Initiator &  initiators that were targeted in a higher volume of MIDs in the past, have higher rates of initiating MIDs.\\
            %$^{*}$ In Degree Target &  targets that were targeted in a higher volume of MIDs in the past, have higher rates of being targeted in MIDs.\\
            %$^{*}$ Out Degree Target &   targets that initiated a higher volume of MIDs in the past, have higher rates of being targeted in MIDs.\\
            $^{*}$ Cyclic closure (itp)&  dyads with a higher number of in-coming two-paths between them, have higher rates of being involved in MIDs.\\
            $^{*}$ Transitive closure (otp) & dyads with a higher number of out-going two-paths, have higher rates of being involved in MIDs.\\
            %$^{*}$ ISP & dyads with a higher number of in-coming shared partners, have higher rates of being involved in MIDs.\\
            %$^{*}$ OSP & dyads with a higher number of out-going shared partners, have higher rates of being involved in MIDs.\\
            $^{*\dagger}$ Major Power & dyads with at least major power actor, have higher rates of being involved in MIDs.\\
            $^{*\dagger}$ Formal Alliance &  dyads involved in a defense alliance have a higher rate of being involved in an MID.\\
            $^{*\dagger}$ Contiguity & dyads with targets and initiators that share a land or water border less than 400 miles, have higher rates of being involved in an MID. \\
            $^{*\dagger}$ Cinc Initator & initiators with a high cinc capabilities score, have higher rates of being involved in MIDs.\\
            $^{*\dagger}$ Cinc Target & targets with a high cinc capabilities score, have higher rates of being involved in MIDs.\\
            $^{*\dagger}$ Log cinc ratio & dyads with a high log cinc capabilities ratio, have higher rates of being involved in MIDs.\\           
            $^{*\dagger}$ Democracy Initiator & initiators with a higher democracy score, have higher rates of being involved in MIDs.\\
            $^{*\dagger}$ Democracy Target & targets with a higher democracy score, have higher rates of being involved in MIDs.\\
            $^{*\dagger}$ Democracy Heterophily & dyads with a high absolute difference in their democracy scores, have higher rates of being involved in MIDs .\\
             $^{*\dagger}$ Trade & dyads with a higher trade flow, have higher rates of being involved in MIDs.\\
        \bottomrule
    \end{tabular}
    \begin{flushleft}
    Legend: \(*\): latent class model variable, \(\dagger\): concomitant model variable
  \end{flushleft}
    \caption{Model statistics and the interpretation of a positive effect of the corresponding statistic.}
    \label{tab:stats}
\end{table}

\subsection{Model specification}
For this application, we model the sequence of MID events that occurred in the period 1946-2012. We utilize the data from 1914-1946 to train the endogenous statistics accounting for the historic MIDs that occurred in this earlier period. These endogenous statistics are updated in the period of 1946-2012 during the observational period. Only states are included that have been involved in at least four MIDs. This results in a relational event history sequence of 2887 events among 113 countries. Table \ref{tab:stats} provides an overview of the endogenous network statistics and the exogenous network statistics. The statistics denoted with a $\dagger$ are also included in the concomitant model. 

Regarding the endogenous stats, we include inertia to account for the tendency of states to keep initiating disputes towards states that they have previously initiated a dispute with. Reciprocity is included to capture the tendencies of states to reciprocate to previously initiated disputes. We normalize the inertia statistic by the out-degree of the initiator state, the reciprocity statistic by the in-degree of the target state. %Moreover, the degree-based statistics such as \textit{out-/in-degree} are also included and are normalized by the total number of observed events. 
The computation of the rrank or recency statistics follows that of \cite{butts_2008_RelationalEventFramework}. For instance, recent target persistence for dyad $(i,j)$ is equal to the inverse of the rank of receiver $j$ among the actors to which sender $i$ has most recently sent past events. Therefore, the most recent receiver corresponds to a statistic of 1, the 2nd most recent to 1/2 and so forth. Additionally, we include two participating shift statistics \citep{gibson_2003_ParticipationShiftsOrder} to capture the diffusion of conflicts beyond the dyad under consideration. For instance, displaced conflicts utilize the p-shift ab-by statistic which captures the tendency for a state which was previously a receiver of a conflict, to target a third state in the immediate next event. Violent spree which utilizes the p-shift ab-ay statistic, accounts for the tendency of an initiator to attack two different states in subsequent events. Triadic statistics such as in-coming two paths (ITP) which captures the tendency for cyclical conflicts in triads and, out-going two paths (OTP) which reflects on the tendency for transitive closures %, in-coming shared partners (ISP), and out-going shared partners (OSP)
\citep{butts_2008_RelationalEventFramework} are also included in the model to account for triadic connections.

In addition to the endogenous statistics above, we also leverage information about the state and dyadic characteristics. Past literature suggests that ``predominance of power balance'' deters military action \citep{russett_2000_ClashCivilizationsRealism,hegre_2008_GravitatingWarPreponderance}. Including the `cinc score' allows us to analyze power balance dynamics, where nations with comparable capabilities may be deterred from engaging in conflicts due to the potential for mutual losses. We specify log cinc ratio, a dyadic covariate, as the logarithmized ratio of the initiator state's capability score divided by the score of the target in the dyad. In addition, we include covariates for the cinc scores of initiator and targets of the disputes. Furthermore, recognizing the role of formal alliances in shaping international relations, formal alliance is included as a covariate in the model. The statistic is a dyadic time-varying covariate with a value of 1 if two states were in a written formal `Type I' defense alliance at that time point. The Direct Contiguity dataset is utilized to establish which states share land or water borders. A dummy variable contiguity is assigned for every dyad that shares a land or water border less than 400 miles. Previous research has also  shown that democratic states are less likely to be involved in disputes with one-another, \citep{russett_2001_TriangulatingPeaceDemocracy,bremer_1993_DemocracyMilitarizedInterstate,benoit_1996_DemocraciesReallyAre,bueno1999,dixon1994,lake1992,maoz1998,oneal1997,ray1995,rousseau1996,russett1993}, to study the role of democratic values in shaping international relations, we incorporate the democracy scores into our model. Following \citep{oneal_2003_CausesPeaceDemocracy} we specify a state's democracy score by subtracting it's autocracy scale score from it's democracy scale score. We include covariates for the initiator's and targets' democracy scores and an additional dyadic covariate corresponding to the absolute difference (referred to as democracy difference) in our model. The values of the democracy score-based statistics are standardized for each event to be on a comparable scale. Several theories of international relations suggest that increased economic interdependence (often measured by trade flow) has a link with the likelihood of conflict \citep{oneal_1999_AssessingLiberalPeace,barbieri_1999_GlobalizationPeaceAssessing,beck_1998_TakingTimeSeriouslya,mansfield_1994_PowerTradeWar,morrow_1999_HowCouldTrade}. Countries engaged in high levels of trade may incur too much economic risk from military conflicts if the mutually beneficial trade is expected to discontinue after conflict \citep{copeland_1996_EconomicInterdependenceWar}. To investigate the relationship between trade and MIDs, we include a dyadic covariate for trade, which corresponds to the value of the directed flow of imports between two countries in a dyad in million USD. The trade-flow value is standardized per year across all the dyads. Lastly, we include a major power dyadic covariate with a value of 1 if at least one of the two actors in that dyad is a major power. This comprehensive set of predictors considers both geopolitical and power-based factors which are deemed to be important in the analysis of Militarized Interstate Disputes.

For the concomitant model, time-varying (endogenous) statistics cannot be included. In the current analysis, the concomitant model contains the same exogenous statistics as the rate part of the DLC-REM. As certain predictors exhibit temporal variations, such as trade, cinc scores, democracy scores etc., we incorporate them into the concomitant model by aggregating their median values over the years. For the time-varying formal alliance statistic we incorporate it into the concomitant model by assigning a value of 1, if the dyad shared a formal alliance at some time point in the observation period.

\subsection{DLC-REM Fitting I: Identifying the Optimal Number of Classes}
The first step when fitting DLC-REMs is to determine the number of dyadic latent classes for the specified model. For this purpose, the AIC, BIC, and recall are computed for DLC-REMs containing different numbers of dyadic latent classes. These DLC-REMs were fitted using the flexmix package as described in Section \ref{sec:flexmix}.

Table \ref{tab:K_comparison} shows the two information criteria and the predictive performance of dyadic latent class models with $K = 1,~2,~3,~4,~5$, or 6 latent classes. For $K=1$, the analysis is equivalent to a standard REM without latent variables. First, the results shows that the AIC decreases as the number of classes increases, suggesting that a model with $K = 6$ dyadic latent classes is preferred. This shows that the AIC has a tendency to prefer (potentially too) complex models. Second, when using the BIC, a model with $K=4$ dyadic latent classes is preferred. This more parsimonious model is preferred because the BIC has a larger penalty for more complex models. Third, based on the recall metric, which was computed using a threshold percentile of 99\%, the model with $K=4$ dyadic latent classes is also preferred as it shows the best predictive performance. Based on these results and to simplify the interpretation, we opt for the more parsimonious model with $K=4$ dyadic latent classes.

\renewcommand{\arraystretch}{1.4}
\begin{table}[t]
\centering
\begin{tabular}{ c c  c c }
\hline
number of latent classes & AIC & BIC & recall \\
\hline
$K = 1$ & 35174 & 37093 & 0.361 \\
$K = 2$ &  34968 & 36080  & 0.441 \\
$K = 3$ & 34940 &  35447 &  0.504 \\
$K = 4$  & 34899 & \textbf{35257} & \textbf{0.535} \\
$K = 5$  & 34489 & 35388 &  0.497 \\
$K = 6$  & \textbf{34229} & 35310   & 0.509 \\
\hline
\end{tabular}
 \caption{Information Criteria and average predictive performance for latent class models with $K = {1,2,3,4,5,6}$. For $K=1$, a standard relational event model is represented. }
\label{tab:K_comparison}
\end{table}

\subsection{DLC-REM Fitting II: Model Interpretation}

\begin{figure}[t]
    \centering
    \includegraphics[width = 0.8\linewidth]{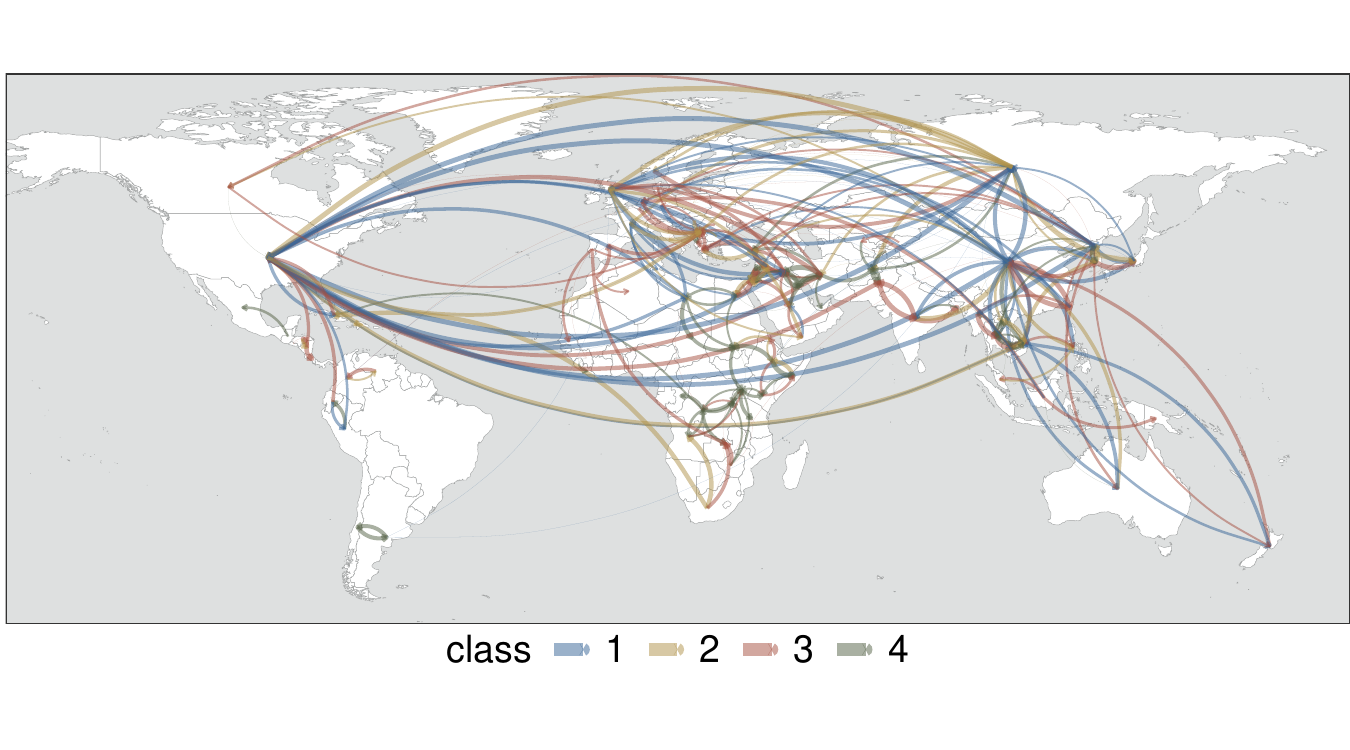}
    \caption{Visualization of the inter-state latent classes for $K$ = 4.}
    \label{fig:map}
\end{figure}

 \begin{table}[htp]
\renewcommand{\arraystretch}{1.4}
\begin{minipage}[pt]{0.95\textwidth}
    \centering
   \fontsize{9pt}{9pt}\selectfont
\setlength\tabcolsep{4pt} % default value: 6pt
%\begin{tabular}{>{\bfseries}l l l l l}
\begin{tabular}{l l l l l }
\toprule
\multicolumn{5}{c}{DLC-REM} \\
\midrule
& class 1 & class 2 & class 3 & class 4\\ 
\midrule
prop. of events & 20\%  & 25\%  & 32\%  & 23\%  \\ 
prop. of dyads & 3\%  & 35\%  & 2\%  & 60\%  \\ 
\midrule
\multicolumn{5}{c}{Rate Model} \\
\midrule
intercept & \textcolor{textcolor1}{-15.193 \ (0.237)} & \textcolor{textcolor1}{-17.079 \ (0.229)} & \textcolor{textcolor1}{-13.132 \ (0.144)} & \textcolor{textcolor1}{-15.072 \ (0.204)} \\
inertia & \textcolor{textcolor1}{-4.045 \ (0.616)} & \textcolor{textcolor2}{0.587 \ (0.348)} & \textcolor{textcolor1}{2.369 \ (0.227)} & \textcolor{textcolor2}{0.199 \ (0.254)} \\
reciprocity & \textcolor{textcolor1}{4.025 \ (0.39)} & \textcolor{textcolor1}{3.279 \ (0.296)} & \textcolor{textcolor2}{0.408 \ (0.266)} & \textcolor{textcolor1}{1.005 \ (0.295)} \\
recent target persistence (rrankSend) & \textcolor{textcolor1}{1.446 \ (0.18)} & \textcolor{textcolor2}{0.307 \ (0.186)} & \textcolor{textcolor1}{1.573 \ (0.145)} & \textcolor{textcolor1}{1.832 \ (0.163)} \\
recent retaliation (rrankReceive) & \textcolor{textcolor1}{0.933 \ (0.187)} & \textcolor{textcolor1}{1.18 \ (0.184)} & \textcolor{textcolor1}{0.816 \ (0.143)} & \textcolor{textcolor1}{0.877 \ (0.176)} \\
displaced conflicts (ps ab-by) & \textcolor{textcolor1}{1.143 \ (0.221)} & \textcolor{textcolor2}{0.2 \ (0.291)} & \textcolor{textcolor1}{-1.736 \ (0.648)} & \textcolor{textcolor1}{1.48 \ (0.263)} \\
violent spree (ps ab-ay) & \textcolor{textcolor1}{1.729 \ (0.158)} & \textcolor{textcolor1}{-1.777 \ (0.316)} & \textcolor{textcolor1}{2.539 \ (0.148)} & \textcolor{textcolor1}{2.001 \ (0.206)} \\
cyclic closure (itp) & \textcolor{textcolor1}{0.02 \ (0.003)} & \textcolor{textcolor2}{0.002 \ (0.005)} & \textcolor{textcolor1}{0.033 \ (0.005)} & \textcolor{textcolor1}{-0.064 \ (0.01)} \\
transitive closure (otp) & \textcolor{textcolor2}{0.007 \ (0.003)} & \textcolor{textcolor1}{0.033 \ (0.005)} & \textcolor{textcolor1}{0.018 \ (0.005)} & \textcolor{textcolor1}{0.085 \ (0.007)} \\
major power & \textcolor{textcolor1}{3.104 \ (0.164)} & \textcolor{textcolor1}{2.7 \ (0.148)} & \textcolor{textcolor1}{1.344 \ (0.169)} & \textcolor{textcolor1}{-2.334 \ (0.419)} \\
formal alliance & \textcolor{textcolor2}{0.226 \ (0.166)} & \textcolor{textcolor1}{0.687 \ (0.147)} & \textcolor{textcolor1}{0.779 \ (0.11)} & \textcolor{textcolor2}{-0.164 \ (0.118)} \\
contiguity & \textcolor{textcolor1}{0.295 \ (0.139)} & \textcolor{textcolor1}{3.536 \ (0.126)} & \textcolor{textcolor1}{1.055 \ (0.103)} & \textcolor{textcolor1}{5.544 \ (0.204)} \\
cinc initiator & \textcolor{textcolor2}{1.424 \ (0.827)} & \textcolor{textcolor2}{0.203 \ (0.93)} & \textcolor{textcolor1}{-3.747 \ (1.384)} & \textcolor{textcolor2}{-6.787 \ (3.715)} \\
cinc target & \textcolor{textcolor1}{2.826 \ (0.86)} & \textcolor{textcolor1}{-6.876 \ (1.507)} & \textcolor{textcolor2}{1.556 \ (1.093)} & \textcolor{textcolor2}{-6.347 \ (3.262)} \\
log cinc ratio & \textcolor{textcolor1}{0.07 \ (0.032)} & \textcolor{textcolor1}{-0.194 \ (0.033)} & \textcolor{textcolor1}{0.409 \ (0.03)} & \textcolor{textcolor1}{0.115 \ (0.036)} \\
democracy initiator & \textcolor{textcolor1}{-1.233 \ (0.198)} & \textcolor{textcolor1}{0.757 \ (0.17)} & \textcolor{textcolor1}{0.428 \ (0.12)} & \textcolor{textcolor1}{-2.519 \ (0.208)} \\
democracy target & \textcolor{textcolor1}{-1.789 \ (0.21)} & \textcolor{textcolor1}{0.811 \ (0.158)} & \textcolor{textcolor2}{0.043 \ (0.122)} & \textcolor{textcolor2}{0.065 \ (0.206)} \\
difference democracy & \textcolor{textcolor1}{3.657 \ (0.216)} & \textcolor{textcolor1}{2.377 \ (0.193)} & \textcolor{textcolor1}{1.183 \ (0.146)} & \textcolor{textcolor1}{-1.528 \ (0.25)} \\
trade & \textcolor{textcolor2}{-0.009 \ (0.013)} & \textcolor{textcolor2}{-0.092 \ (0.05)} & \textcolor{textcolor1}{-0.053 \ (0.027)} & \textcolor{textcolor1}{0.093 \ (0.012)} \\
%%%%%%%%%%%%%%%%%%%%%%%%%%%%%%%%%
\midrule
\multicolumn{5}{c}{Concomitant Model} \\
\midrule %%%%%%%%%%%%%%%%%%%%%%%%%%%%%%%%%
intercept & -1.944 \ (0.077) & -0.442 \ (0.057) & -3.018 \ (0.124) \\ 
major power & 0.255 \ (0.088) & \textcolor{textcolor2}{-0.058 \ (0.088)} & \textcolor{textcolor2}{-0.116 \ (0.138)} \\ 
formal alliance & 0.562 \ (0.087) & -0.402 \ (0.087) & 0.466 \ (0.141) \\ 
contiguity & 1.435 \ (0.114) & \textcolor{textcolor2}{-0.023 \ (0.135)} & 2.111 \ (0.149) \\ 
cinc initiator & {0.854 \ (0.083)} & {0.401 \ (0.0128)} & {-2.812 \ (0.215)} & \\ 
cinc target & {1.288 \ (0.02)} & \textcolor{textcolor2}{-0.228 \ (0.62)} & {0.441 \ (0.12)} & \\ 
log cinc ratio & 0.105 \ (0.0149) & 0.511 \ (0.012) & -0.171 \ (0.024) \\ 
democracy initiator & \textcolor{textcolor2}{0.147 \ (0.084)} & 0.351 \ (0.069) & -0.969 \ (0.148) \\ 
democracy target  & 0.434 \ (0.083) & -0.448 \ (0.07) & \textcolor{textcolor2}{0.06 \ (0.141)} \\ 
difference democracy  & 0.961 \ (0.107) & 0.25 \ (0.092) & 1.828 \ (0.178) \\ 
trade & -0.214 \ (0.06) & -0.186 \ (0.073) & \textcolor{textcolor2}{-0.129 \ (0.095)} \\ 
\bottomrule
\end{tabular}
      \captionof{table}{DLC-REM and corresponding concomitant model estimates for $K$ = 4 with standard errors in brackets. Class 4 is the reference class for the concomitant model.}% Coefficients with $p$ value $<0.05$ are in black font, and coefficients with $p$ value $>0.05$ are in grey font.}
\label{table:coefficients}
\end{minipage}
\end{table}
Table \ref{table:coefficients} reports the estimates and standard errors of the dyadic latent class model for $K=4$ classes. In the corresponding concomitant model, class 4 serves as the reference class because it contains most dyads and thus represents the most typical behavior. The variations in the estimates for various effects across classes suggests that there is indeed heterogeneity in the interaction mechanisms of the dyads. Figure \ref{fig:map} depicts the classification of the dyads on a world map and Table \ref{tab:countries} lists the top 6 most active dyads in each class.

\begin{table}[ht]
\centering
\begin{tabular}{ll}
\toprule
\textbf{Class 1} & \textbf{Class 2} \\
\midrule
China $\rightarrow$ United States of America & Russia $\rightarrow$ United States of America \\
United States of America $\rightarrow$ China & Russia $\rightarrow$ Japan \\
United States of America $\rightarrow$ Russia & Israel $\rightarrow$ Egypt \\
United States of America $\rightarrow$ Iraq & Israel $\rightarrow$ Syria \\
China $\rightarrow$ Russia & India $\rightarrow$ China \\
United Kingdom $\rightarrow$ Iraq & Israel $\rightarrow$ Jordan \\
\midrule
\textbf{Class 3} & \textbf{Class 4} \\
\midrule
Pakistan $\rightarrow$ India & North Korea $\rightarrow$ South Korea \\
Turkey $\rightarrow$ Greece & Afghanistan $\rightarrow$ Pakistan \\
Thailand $\rightarrow$ Cambodia & Ethiopia $\rightarrow$ Somalia \\
Israel $\rightarrow$ Lebanon & Kuwait $\rightarrow$ Iraq \\
Myanmar $\rightarrow$ Thailand & China $\rightarrow$ Vietnam \\
Turkey $\rightarrow$ Iraq & Iraq $\rightarrow$ Kuwait \\
\bottomrule
\end{tabular}
\caption{List of 6 most active dyads per class }
\label{tab:countries}
\end{table}

%Class 1 represents 20\% of events and 3\% of dyads, indicating a less common but distinct type of interaction. Class 2 accounts for 25\% of events and a large proportion (35\%) of dyads, indicating these interactions are frequent and involve many dyads. Class 3 has the highest proportion of events (32\%) but is rare in terms of dyads (2\%), indicating that a few specific dyads have very high interaction rates. Class 4, the reference class with 23\% of events and the majority (60\%) of dyads, which may represent the most typical or general type of dyadic interaction.
Table \ref{table:coefficients} shows that the observed events are roughly equally spread across the four classes but the sizes of the dyadic classes (in terms of the number of dyads) highly varies. Classes 1 and 3 are rather small (consisting of 3\% and 2\% dyads) and classes 2 and 4 are rather large (consisting of 35\% and 60\% dyads). These results indicate that relatively active dyads (in classes 1 and 3) tend to show different interaction behavior than relatively inactive dyads based on the information contained in the predictor statistics.

The results of the dyadic latent class relational event model (DLC-REM) show nuanced patterns of militarized disputes across four classes of dyads, each with unique characteristics. The concomitant model coefficients describe how various factors predict the likelihood of dyads belonging to each class compared to the reference class (class 4). Class 1 primarily comprises dyads involving large, powerful countries, often characterized by formal alliances, shared borders, and differences in democratic governance. The frequent interactions between major powers in this class, like China and USA or USA and Russia  (Table \ref{tab:countries}) indicate a tendency toward conflict, where reciprocal and retaliatory behaviors play a significant role (Table \ref{table:coefficients}). These dyads display high reciprocity and recent reciprocation effects, indicating a propensity for immediate response to actions. The presence of major powers, significant democratic differences and material capabilities (as measured by the cinc score) of the target increases the likelihood of MIDs.

The dyads in Class 3, by contrast, are generally composed of less democratic initiators with lower capabilities and large democratic differences with their targets, alongside shared borders. Active dyads in this class, such as Pakistan-India and Turkey-Greece, align with these attributes (Table \ref{tab:countries}). Here, inertia and recent target persistence effects suggest that past disputes increase the likelihood of future MID events. Additionally, violent spree effects indicate that dyads in Class 3 are prone to aggressive behaviors. The presence of democratic initiators with lower capabilities have a positive effect on event rate suggesting that weaker states may initiate conflicts.

The concomitant model coefficients in Class 2 are less pronounced without strong indicators like formal alliances or democratic differences seen in Classes 1 and 3. This can also be explained by this class containing a large number of dyads (35\%). Major powers still appear in this class, but the absence of formal alliances stands out as a key feature, aligning with the presence of non-allied dyads such as Russia-USA, India-China and Israel-Egypt. Compared to Classes 1 and 3, Class 2 has lower levels of conflict initiation and escalation, with non-significant effect for  displaced conflicts and inertia effect, suggesting these dyads are less prone to conflict escalation.

In Class 4, where a majority of dyads are grouped, event rates are primarily influenced by geographic proximity and recent reciprocation. Contiguity has a strong positive effect, suggesting that neighboring countries in this class are more prone to MIDs. This can also be seen from the most active dyads belonging to this class (Table \ref{tab:countries}).

 The impact of trade on conflict has been a topic of longstanding debate among international relations scholars. While some argue that economic interdependence through trade increases interstate disputes, others suggest it fosters peace due to mutual benefits and acts as a deterrent to disputes \citep{barbieri_1996_EconomicInterdependencePath,copeland_1996_EconomicInterdependenceWar,oneal_1999_AssessingLiberalPeace}. The results here suggest a largely negligible influence of trade on conflict. In Classes 1 and 3, trade slightly reduces the disputes rate, suggesting a limited pacifying effect, but overall, trade effects are minimal and inconclusive across the model. This finding aligns with the mixed literature on trade, suggesting that trade might not significantly deter disputes and could correlate with higher event rates. 

The impact of alliances has been another important topic in the literature  \citep{oneal_2003_CausesPeaceDemocracy, benson_2011_UnpackingAlliancesDeterrent}. The results of the fitted dyadic latent class relational event model also reveals intricate dynamics in militarized dispute events among dyads. The positive and significant formal alliance effect in  Classes 1 and 3 may be surprising as this implies that dyad states with a formal alliance are more likely to be involved in an MID when the endogenous effects and other covariates are held constant. This positive effect of alliances has been observed in previous studies, suggesting that alliances can sometimes escalate conflicts rather than deter them \citep{oneal_2003_CausesPeaceDemocracy, benson_2011_UnpackingAlliancesDeterrent}. Thus, the role of formal alliances in conflict dynamics is complex and context-dependent.

Finally, the impact of democracy on militarized interstate disputes (MIDs) has also been a longstanding topic in international relations research, with a prevailing theory known as the ``democratic peace" suggesting that democracies are less likely to engage in conflict, particularly with one another. Results from the DLC-REM model provide a nuanced view, showing that democracy influences disputes differently across dyadic classes. For instance, in Class 1, where major power dynamics are significant, higher democratic scores in either the initiator or target generally reduce disputes rates, consistent with democratic peace theory. Whereas in Classes 2 and  3, democratic countries as either initiator or target are associated with a higher likelihood of disputes. Additionally, the effect for democracy differences within dyads is associated with an increased likelihood of conflict across the Classes 1,2, and 3 indicating that sharp disparities in governance models can exacerbate tensions and lead to more frequent disputes.

DLC-REM reveals distinct patterns of militarized disputes across different dyadic classes, where major power status, alliances, democratic differences, and geographical proximity are prominent predictors of conflict behavior.

\section{Discussion}
\label{sec6}
In this article we proposed the dyadic latent class relational event model (DLC-REM) for capturing unobserved heterogeneity regarding social interaction behavior between actors in a relational event network. This model addresses the important limitation of the standard relational event model that assumes all dyads to have the same tendency to interact with one another as a function of their exogenous and endogenous statistics (predictors). Our model provides deeper, more nuanced insights into the heterogeneity of interaction mechanisms between dyads in a network. We employed simulations to illustrate the efficacy of dyadic latent classes in capturing the subtleties in identifying dyadic latent classes and compared it with a stochastic block model where actors, rather than dyads, are categorized in latent classes. This illustrated that the DLC-REM is more flexible than the SB-REM by allowing more complex latent structures with the same number of unknown parameters (i.e., total number of coefficients across classes). The simulations demonstrated the predictive ability and generalizability of our model particularly as an alternative to blockmodels popularly used in the social network modelling literature. Based on these results we argue that the proposed DLC-REM may be preferred over stochastic block modeling approaches to explain and predict relational event data in the case of complex social structures.

The application on militarized dispute events using DLC-REM with four classes provided key insights into the complex interplay of factors influencing conflict dynamics. The findings highlight that trade and formal alliances have nuanced roles, sometimes deterring and other times escalating conflicts, depending on the context. Similarly, the impact of democracy varies, generally supporting the democratic peace theory but also showing that democratic states can engage in disputes under certain conditions. These results highlight the importance of considering heterogeneity of endogenous and exogenous factors, as well as the specific characteristics of dyads, to understand the multifaceted nature of international conflicts.

In comparison to prior work to model unobserved heterogeneity \citep{juozaitiene_2022_NodalHeterogeneityMayb,butts_2008_RelationalEventFramework,uzaheta_2023_RandomEffectsDynamica,mulder2024latent}, the marginal product mixture model (MPMM) of \cite{dubois_2010_ModelingRelationalEvents} is perhaps closest to the approach detailed in this article, although there are certain key methodological and conceptual differences. Our approach includes dyadic endogenous and exogenous covariates in the determination of the classes and class-specific effects while the MPMM model clusters the dyads into classes based on their interaction frequency. This not only makes our approach more general but also changes the interpretation of the classification. While a frequency-based classification implies that dyads that belong to the same class interact with others at a similar frequency, a statistics-based classification implies that dyads grouped in the same class have similarity in their interaction mechanisms. Further, the MPMM model proposed does not model the waiting time between events which is a crucial feature for modelling relational events. In addition, our approach allows for incorporating a concomitant model that enhances our understanding of the factors that influence latent class membership.

Instead of estimating the DLC-REM with the concomitant model jointly, there are also approaches in which the concomitant model is estimated in a second step after fitting the latent class model \citep{vermunt_2017_LatentClassModeling}. After identifying the latent classes in the first step, the class memberships of the observations (dyads in our case) are assumed known and then a multinomial model is fitted in a separate second step. This two-step approach can be computationally faster, and therefore may be preferred in applied work. A potential limitation of this method however may be that any uncertainty regarding class assignment is ignored in the second step. A thorough comparison between these two approaches for the DLC-REM would be useful for future work.

DLC-REMs allow for handling simultaneous events in a natural way for relational event models. For instance, in scenarios involving international disputes, simultaneous conflicts might unfold across different locations or involve multiple parties, each acting independently yet being interconnected. This makes it difficult to discern the precise sequence of actions without more granular data. Digital news media also often log relational events by merely noting the date, neglecting the specific times or the sequence of events, a limitation highlighted in the research by \cite{brandes_2009_NetworksEvolvingStep}. This can also apply to data from proximity sensors, which might record times in set time intervals. Although the standard Relational Event Model (REM) does not support concurrent events, by adapting a Poisson process model for relational event histories, it is possible to handle this scenario effectively without ad hoc solutions such as adding random noise to event times that occurred simultaneously or arbitrarily splitting events across the interval in an arbitrary order, which could introduce potential biases.

DLC-REMs allow for various options for handling latent class assignment. One possibility is to assume that dyads in both directions (e.g., actor A towards actor B and actor B towards actor A) share the same dyadic latent class. Another option is to assign non-observed dyads to a single class. This bears some similarity to SB-REM, which also imposes a specific latent structure on dyads. The choice of how to handle latent class assignment introduces nuances into the model's interpretation and should be considered carefully based on the underlying dynamics of the relational events under investigation. In the context of international militarized disputes, however, we did not impose any constraints. We allow the relation from A to B to be entirely different from that from B to A. For instance, a major power may have very different considerations and triggers when engaging in conflict with a weaker target than the other way around. Our dyadic latent class model allows for asymmetric class assignment such that the dyad A to B may belong to a different class than the B to A dyad. However, it is straightforward to constrain the model to allow for symmetric class assignment if the research question requires it. Such a DLC-REM could also be fit in the same manner as discussed in this paper.

Furthermore, the application on the interstate disputes has certain limitations that should be considered in its interpretation and generalization. First, while the model effectively captures the occurrence of MID events, it does not explicitly account for the durations of these events. Incorporating the duration of MIDs could provide a more nuanced understanding of their dynamics. Second, the current model primarily focuses on dyadic ties, and there is a limitation in not accounting for events involving multiple actors simultaneously. Incorporating a framework to model hyper-events, where interactions involve more than two actors, could enhance the insights gained. We leave this for future research.

\section*{Acknowledgements}
The authors wish to thank Jeroen Vermunt for helpful suggestions and feedback. This research was supported by a Vidi grant from the Netherlands Organization for Scientific Research (NWO; 452-17-006) awarded to the third author.

\newpage
\bibliography{references}

\begin{appendices}
    \section{R code of simulations for Section \ref{sim}}
    \label{appendix:a}
    Simulating relational event sequences with structure depicted in Figure \ref{plot:sim}(\subref{plot:sim_gen})
\begin{lstlisting}
library(remulate)

#Source functions
source("./functions.R")

# Define number of simulations to repeat
runs = 100
# Define the effects for which to generate the event history
effects = c("baseline","inertia","reciprocity","psABBA","psABBY","psABAY")

# Length of the effects vector, used to dimension arrays
P <- length(effects)

# Set up simulation parameters
K <- 4  # Number of classes
M <- 2000  # Number of events per sequence
N <- 10  # Number of actors

# Seed for reproducibility
set.seed(123)

# Initialize a 3D array to store beta coefficients for each effect and pair of actors
beta_mat_gen <- array(0, c(N, N, P))
dimnames(beta_mat_gen)[[3]] <- effects

# Set values for the 'baseline' effect in the beta matrix
beta_mat_gen[,,"baseline"] = -11
beta_mat_gen[1:4,1:4,"baseline"] = -2
beta_mat_gen[2:10,8:10,"baseline"] = -5
beta_mat_gen[6:9,2:3,"baseline"] = -3
beta_mat_gen[8:9,4:6,"baseline"] = -3

# Set the values for the 'inertia' effect
beta_mat_gen[,,"inertia"] = -0.2
beta_mat_gen[1:4,1:4,"inertia"] = 0.1
beta_mat_gen[2:10,8:10,"inertia"] = 0.6
beta_mat_gen[6:9,2:3,"inertia"] = 0.3
beta_mat_gen[8:9,4:6,"inertia"] = 0.3

# Set the values for 'reciprocity'
beta_mat_gen[,,"reciprocity"] = -0.3
beta_mat_gen[1:4,1:4,"reciprocity"] = 0.05
beta_mat_gen[2:10,8:10,"reciprocity"] = 0.1
beta_mat_gen[6:9,2:3,"reciprocity"] = 0.2
beta_mat_gen[8:9,4:6,"reciprocity"] = 0.2

# Generate random effects for 'psABBA', 'psABBY', 'psABAY' with different ranges
eff = round(sort(runif(4,1,3)),2)
beta_mat_gen[,,"psABBA"] = eff[1]
beta_mat_gen[1:4,1:4,"psABBA"] = eff[2]
beta_mat_gen[6:9,2:3,"psABBA"] = eff[3]
beta_mat_gen[8:9,4:6,"psABBA"] = eff[3]
beta_mat_gen[2:10,8:10,"psABBA"] = eff[4]

eff = round(sort(runif(4,1,2)),2)
beta_mat_gen[,,"psABBY"] = eff[1]
beta_mat_gen[1:4,1:4,"psABBY"] = eff[2]
beta_mat_gen[6:9,2:3,"psABBY"] = eff[3]
beta_mat_gen[8:9,4:6,"psABBY"] = eff[3]
beta_mat_gen[2:10,8:10,"psABBY"] = eff[4]

eff = round(sort(runif(4,1,1)),2)
beta_mat_gen[,,"psABAY"] = eff[1]
beta_mat_gen[1:4,1:4,"psABAY"] = eff[2]
beta_mat_gen[6:9,2:3,"psABAY"] = eff[3]
beta_mat_gen[8:9,4:6,"psABAY"] = eff[3]
beta_mat_gen[2:10,8:10,"psABAY"] = eff[4]


# Generate the formula object for remulate
remulate_gen <- beta_mat_to_remulate(beta_mat_gen, effects, scaling="prop")
cov <- remulate_gen$cov

# Define a suffix for file naming
suffix = "_aug"

# Loop to generate data and save files
for(r in 1:runs){
    dat_gen <- remulate::remulateTie(remulate_gen$form, 1:N, 100000, M)
    dat_gen$probs = NULL
    save(dat_gen, file=paste0(data_dir, "dat_gen_K=", K, "_r=", r, suffix, ".rdata"))
}

# Save a generated object for visualization
gen.obj <- list(
    beta_mat_gen = beta_mat_gen,
    N = N,
    M = M,
    cov = cov
)

save(gen.obj, file=paste0(data_dir, "gen.obj_K=", K, suffix, ".rdata"))

# Generate and save plots for each effect
glist <- plot_params(model = "beta_mat", r = 1, K = K, data_gen.obj = gen.obj, effects = effects, fit_obj = beta_mat_gen, row_order = N:1, col_order = 1:N, tile_label = FALSE, palette = "paired")

glist[[1]] + ggtitle("")
g <- glist[[1]] + ggtitle("")
ggsave(plot=g, filename = paste0(plots_dir, "gen_baseline_r_", r, suffix, ".pdf"), width = 5, height = 5, device='pdf', dpi = 600)

g <- glist[[2]]  + ggtitle("")
ggsave(plot=g, filename = paste0(plots_dir, "gen_inertia_r_", r, suffix, ".pdf"), width = 5, height = 5, device='pdf', dpi = 600)



\end{lstlisting}  
    
    Helper functions required:
\begin{lstlisting}

#'
#'@param beta_mat NxNxP
#'@param effects vector of effect names to use in remulate in order corresponding to P elements in beta_mat
#'@return remulate formula
#'@return list containing form , a formula object for remulate and cov object for remulate input
beta_mat_to_remulate <- function(beta_mat,effects,scaling = c("raw", "std", "prop")){
    brem_effects <- c("baseline","psABBA","psABBY","psABXA","psABXB","psABAY",
                      "outdegreeSender","outdegreeReceiver","indegreeSender",
                      "indegreeReceiver","inertia","changepoint_count","reciprocity")
    
    if(any(!effects %in% brem_effects)){
        stop("effect specified is not supported")
    }
    if(dim(beta_mat)[3] != length(effects)){
        stop("dim of beta_mat doesnt match effects argument")
    }
    
    N <- dim(beta_mat)[1]
    
    rs <- data.frame(sender=rep(1:N,N),receiver=rep(1:N,each=N))
    rs <- rs[rs[,1] != rs[,2],]
    
    cov <- data.frame(matrix(0,nrow(rs),2+length(effects)))
    cov[,1:2] <- rs
    colnames(cov) <- c("sender","receiver",effects)
    
    cov[,3:(2+length(effects))] <-  t(apply(cov[,1:2],1,function(x){
        beta_mat[x[1],x[2],]
    }))
    
    #remulate formula
    if(any(effects=="baseline")){
        b_ind = which(effects=="baseline")
        form <- paste0("~ remulate::baseline(0) + remulate::dyad(0,'baseline',cov) + ", paste(paste0("remulate::",effects[-b_ind],"(0,scaling='",scaling,"') + remulate::dyad(0,'",effects[-b_ind],"', cov)") , collapse = " + ") ," + ", paste(paste0("remulate::interact(1, c(",( 2 * 1:length(effects) - 1),",",(2*(1:length(effects))),"))") , collapse=" + ")) 
    }else{
        form <- paste0("~", paste(paste0("remulate::",effects,"(0,scaling='",scaling,"') + remulate::dyad(0,'",effects,"', cov)") , collapse = " + ") ," + ", paste(paste0("remulate::interact(1, c(",( 2 * 1:length(effects) - 1),",",(2*(1:length(effects))),"))") , collapse=" + "))    
    }
    return(list(form= as.formula(form),cov=cov))
    
}


#' Function to plot params
#' @param model string specifying the type of fit_obj if(model=="beta_mat")
#' then the beta mat is used directly instead of computing it from the fit_obj
#' 
#' @param row_order order of rows in the heatmap
#' @param col_order order of cols in the heatmap
#' @param effects order fo effects in the grid 
#' @param palette RColorBrewer palette
#' 
plot_params <- function(model=c("lc","sbm","beta_mat"),data_gen.obj,r,K,effects,fit_obj=NULL,palette = 'paired',palette_direction = 1,row_order = NULL, col_order = NULL,tile_label = TRUE){
    if(is.null(fit_obj)){
        #plotting generating params
        if(model=="sbm"){
            beta_mat_m <-  brem_params_to_beta_mat(data_gen.obj$brem_model,1:data_gen.obj$N)
            
        }else if(model=="lc"){
            beta_mat_m <-  flexmix_params_to_beta_mat(data_gen.obj$lc_model,1:data_gen.obj$N,data_gen.obj$M)    
        }    
    }else{
        #plotting fitted params
        if(model=="sbm"){
            beta_mat_m <-  brem_params_to_beta_mat(fit_obj,1:data_gen.obj$N)
            
        }else if(model=="lc"){
            beta_mat_m <- flexmix_params_to_beta_mat(fit_obj,1:data_gen.obj$N,data_gen.obj$M)
        }
    }
    if(model=="beta_mat"){
        beta_mat_m = fit_obj
    }
    if( "(Intercept)" %in% dimnames(beta_mat_m)[[3]]){
        dimnames(beta_mat_m)[[3]][which(dimnames(beta_mat_m)[[3]] == "(Intercept)")] = "baseline"
    }
    #re-order the effects
    beta_mat_m = beta_mat_m[,,effects]
    
    remulate_m <- beta_mat_to_remulate(beta_mat_m,effects,scaling="prop")
    g.cov <- remulate_m$cov  #gen cov
    
    g.cov[,3:ncol(g.cov)] <- round(g.cov[,3:ncol(g.cov)],2)
    for(eff in colnames(g.cov)){
        g.cov[[eff]] <- factor(g.cov[[eff]])
    }
    if(!is.null(row_order) & !is.null(col_order)){
        g.cov <- g.cov %>%
            mutate(Sender = factor(sender, levels = as.character(rev(col_order))),
                   Receiver = factor(receiver, levels = as.character(rev(row_order))))
    }else{
        g.cov <- g.cov %>%
            mutate(Sender = factor(sender, levels = as.character(rev(1:data_gen.obj$N))),
                   Receiver = factor(receiver, levels = as.character(rev(1:data_gen.obj$N))))
    }
    
    #rev(levels(receiver))
    glist <- list()
    if(tile_label){
        for(eff in effects){
            g <- ggplot(g.cov, aes_string(x="Receiver", y="Sender", fill= eff)) +
                geom_tile(color = "#666666",width=0.99, height=0.99) +
                ggtitle(eff)+
                geom_text(aes_string(label = eff), color = "black", size = 5) +
                theme(legend.position = "none")+
                coord_fixed()+
                theme(axis.title=element_text(size=20))+ # axis title size
                theme(axis.text.x = element_text(angle = 90, vjust = 0.5, hjust=1,size = 16))+
                theme(axis.text.y= element_text(size = 16))+
                scale_fill_brewer(palette = palette,direction = palette_direction)
            glist[[eff]] <- g
        }    
    }else{
        for(eff in effects){
            g <- ggplot(g.cov, aes_string(x="Receiver", y="Sender", fill= eff)) +
                geom_tile(color = "#666666",width=0.99, height=0.99) +
                ggtitle(eff)+
                #geom_text(aes_string(label = eff), color = "black", size = 2) +
                theme(legend.position = "none")+
                coord_fixed()+
                #scale_x_discrete(labels = as.character(1:data_gen.obj$N))+
                #scale_y_discrete(labels = as.character(1:data_gen.obj$N))+
                theme(axis.title=element_text(size=20))+ # axis title size
                theme(axis.text.x = element_text(angle = 90, vjust = 0.5, hjust=1,size = 16))+
                theme(axis.text.y= element_text(size = 16))+
                scale_fill_brewer(palette = palette,direction = palette_direction)
            glist[[eff]] <- g
        }
    }
    
    return(glist)
}
\end{lstlisting}
    \section{Pre-processing function for fitting DLC-REM}
    \label{appendix:b}
\begin{lstlisting}
# Prepare relational event data for GLM analysis
# 
# This function formats an edge list and associated statistics for analysis
# in a generalized linear model (GLM), for relational event network data.
# It creates a data frame suitable for GLM analysis with a binary response 
# variable indicating the (non) occurrence of an event between each possible
# dyad in the riskset at each time point.
#
# @param edgelist A data frame with columns time, sender and  receiver of observed events.
# @param statistics An array of statistics associated with each event in the edgelist.
# @param evls A matrix with columns dyad id and time of observed events.
# @param N Integer, the number of actors in the network.
# @param rs A matrix representing the riskset i.e the sender and receiver ids for dyads corresponding to the statistics array and evls.
# @param mult Logical, indicating whether there are multiple events per time point (default FALSE).
# @return A list containing a data frame ready for GLM and a vector of log interevent times.
prepare_reh_glm <- function(edgelist, statistics, evls, N, rs, mult = FALSE) {
    
    # Calculate the total number of possible directed dyads (excluding self-ties)
    ties <- N * (N - 1)
    # Total number of events
    E <- dim(statistics)[1]  
    
    # Process time data and create log time differences for the offset
    if (mult == TRUE) {
        edgelist <- edgelist[order(edgelist$time),]
        keep_stat_rows = !duplicated(edgelist$time, fromLast = TRUE)
        unique_times = edgelist$time[keep_stat_rows]
        log_intevent <- log(unique_times - c(0, unique_times[-E]))
    } else {
        edgelist <- edgelist[order(edgelist$time),]
        log_intevent <- log(edgelist$time - c(0, edgelist$time[-E]))
    }
    
    # Construct the statistics stack for each event
    statStack <- do.call(rbind, lapply(1:E, function(e) {
        statistics[e, , ]
    }))
    stat_glm <- as.data.frame(statStack)
    stat_glm$log_intevent <- rep(log_intevent, each = ties)
    
    # Create the binary response variable
    stat_glm$obs <- unlist(lapply(1:E, function(e) {
        indic <- rep(0, ties)
        if (mult) {
            obs_dyads = evls[which(evls[, 2] == unique_times[e]), 1]
        } else {
            obs_dyads = evls[e, 1]
        }
        indic[obs_dyads] <- 1
        indic
    }))
    
    # Include dyad index for each row
    stat_glm$tie <- rep(1:ties, E)
    stat_glm_tiesorted <- stat_glm[order(stat_glm$tie),]
    
    return(list(stat_glm = stat_glm_tiesorted, log_intevent = log_intevent))
}
\end{lstlisting}  
\end{appendices}

\end{document}